\documentclass[12pt]{article}
\usepackage{epsfig}
\usepackage{amsmath}
\usepackage{amsfonts}

\def\pslash{\hbox{/\kern-.5800em$p$}}

\def\gappeq{\mathrel{\rlap {\raise.5ex\hbox{$>$}}
{\lower.5ex\hbox{$\sim$}}}}
\def\lappeq{\mathrel{\rlap{\raise.5ex\hbox{$<$}}
{\lower.5ex\hbox{$\sim$}}}}

\begin{document}
\pagestyle{empty}
\begin{flushright}
UMN-TH-2408/05\\
July 2005
\end{flushright}
\vspace*{5mm}

\begin{center}
{\Large\bf Emergent Gravity from a Mass Deformation \\ in Warped Spacetime}
\vspace{1.0cm}

{\sc Tony Gherghetta$^a$}\footnote{E-mail:  tgher@physics.umn.edu}, 
{\sc Marco Peloso$^a$}\footnote{E-mail:  peloso@physics.umn.edu}
{\small and} 
{\sc Erich Poppitz$^b$}\footnote{E-mail:  poppitz@physics.utoronto.ca}
\\
\vspace{.5cm}
{\it\small {$^a$School of Physics and Astronomy\\
University of Minnesota\\
Minneapolis, MN 55455, USA}}\\
\vspace{.5cm}
{\it\small {$^b$Department of Physics\\
University of Toronto\\
Toronto, ON M5S 1A7, Canada}}\\
\end{center}

\vspace{1cm}
\begin{abstract}
We consider a deformation of five-dimensional warped gravity 
with bulk and boundary mass terms to quadratic order in the action.
We show that massless zero modes occur for  special choices of the masses. 
The tensor zero mode is a smooth deformation of 
the Randall-Sundrum graviton wavefunction and can be localized anywhere 
in the bulk. There is also a vector zero mode with similar localization 
properties, which is decoupled from conserved sources at tree level. 
Interestingly, there are no scalar modes, and the model is ghost-free at 
the linearized level. When the tensor zero mode is localized 
near the IR brane, the dual interpretation is a composite  graviton
describing an emergent (induced) theory of gravity at the IR scale. In this 
case Newton's law of gravity changes to a new power law below the 
millimeter scale, with an exponent that can even be irrational.

\end{abstract}

\vfill
\begin{flushleft}
\end{flushleft}
\eject
\pagestyle{empty}
\setcounter{page}{1}
\setcounter{footnote}{0}
\pagestyle{plain}

\section{Introduction}

It is a striking fact that four-dimensional (4D) 
gravity can be localized in AdS$_5$ by tuning 
the bulk and brane cosmological constants~\cite{rs}. It is even more 
remarkable  that this five-dimensional (5D) model has a dual 4D 
interpretation via the AdS/CFT 
correspondence~\cite{adscft1}-\cite{pheno3}.
The gauge-gravity duality has made the warped gravity setup 
an attractive setting for studying aspects of strongly coupled gauge 
dynamics, from both the formal and phenomenological perspective. 
A feature particularly useful for low-energy model building is that 
 the nonfactorizable geometry localizes not only gravity, but also fields of 
different spin, such as scalars and fermions~\cite{grossneub,gp1}.

An important difference of  scalar and fermion localization 
from gravity is that fermion and scalar zero modes can be localized anywhere 
in the 5D bulk. This unrestricted localization 
is achieved by introducing bulk and boundary masses, with the degree of 
localization directly depending on the bulk mass parameter~\cite{gp1}. This 
has been particularly fruitful for model building, where models of UV/IR brane-localized fields, 
corresponding to hybrid theories of elementary and composite 
particles in the 4D dual, have opened new phenomenological
possibilities for the supersymmetric standard model~\cite{gp2,gp4} and the 
Higgs sector~\cite{cnp,acp}. In fact, this unrestricted localization can also 
occur for gauge fields. While it is well known that gauge bosons are not 
localized in the bulk~\cite{gaugeb1,gaugeb2}, it is possible to tune bulk and 
boundary mass terms so that a U(1) zero mode can be localized 
anywhere~\cite{gn}.

The graviton appears to be different, but it is natural to ask whether 
gravity can similarly be localized anywhere in the bulk. The AdS/CFT 
correspondence provides our  primary motivation for studying the 
delocalization of the graviton in the Randall-Sundrum (RS) scenario. If the 
graviton zero mode could be localized on the IR brane, this would suggest that in the dual 
theory the graviton is a composite CFT state whereby dynamical gravity 
only emerges in the infrared. The dual UV theory would then be a pure 
gauge theory with no propagating massless spin-2 particle. It has been 
suggested that in an ``emerging" (``fat'') gravity model, the graviton 
compositeness may alleviate the UV sensitivity of the cosmological 
constant~\cite{fatgrav}.  One may worry that delocalization 
comes at the price of losing general covariance in the bulk, but as we 
will see, this does not preclude the existence of a massless tensor mode. 
Furthermore, the Weinberg-Witten theorem is avoided if gravity 
is induced at the quantum level~\cite{ww}. Thus, there does not appear to 
be any obstacle to smoothly deforming the graviton wave function away from 
the UV brane, as can be done for other bulk fields, and one hopes that a 
similar mechanism can be implemented for gravity.

In this paper, we study a modification of the linear Einstein equations and show that, 
indeed, the graviton can be localized on the IR brane. Much like for 
bulk scalar and fermion fields, our modification requires the introduction 
of bulk and boundary mass terms for the metric perturbations. However adding
a graviton mass is technically more involved compared to the case of
lower spin fields. For instance, already in 4D a Fierz--Pauli~\cite{fipa}
mass term for the graviton leads to the well known vDVZ discontinuity at the
linearized level~\cite{vdvz} (while other choices lead to ghosts).
This is due to the fact that the longitudinal polarization of a massive graviton
remains coupled to matter
in the limit of zero graviton mass. In the warped 5D case, the extra
polarizations of the graviton can be systematically studied 
by exploiting the symmetries of the 4D Minkowski slices. This leads to 
the definition of tensor, vector, and scalar fluctuations with respect to 
the 4D Poincar{\'e} transformations. In this way, we show that by a 
special choice of the bulk and boundary mass terms there exist 
tensor and vector zero modes that 
can be localized anywhere in the bulk. A priori, one should also expect 
scalar modes but, remarkably, they are absent in our model.
Thus, up to quadratic order in the action, we will see that our model 
is ghost-free. This is in contrast to many modifications of gravity
where ghost perturbations typically arise in the scalar sector. 
In particular, a previous study of massive gravity in warped geometry, 
which only introduced boundary mass terms,   
concluded that ghosts were present in the spectrum~\cite{cggp}.
\smallskip

{\flushleft B}elow, we present the structure and summarize  the main 
results of this paper: 
\begin{itemize}
\item
We begin, in Section 2, with a study of the bulk equations of motion for 
the graviton  perturbation with a bulk mass term added. The solutions of 
the  linearized equations of motion for the metric perturbations  lead to 
massless and massive tensor, vector, and scalar modes. These are 
subject to boundary conditions on the two branes, which we discuss next. 
\item Boundary mass terms are introduced
and the junction conditions are derived, in Section 3, for all the metric 
perturbations. Some details are presented in the Appendices; in Appendix A 
we present the bulk gravity action to quadratic order. 
In particular, in Section 3 we also find that a massless tensor and vector
mode is consistent with the boundary conditions and that the model is 
ghost-free. For any allowed value of the bulk mass parameter, there exist two 
possible choices of boundary terms  admitting a zero mode 
(see Fig.~1).~\footnote{Curiously, for zero bulk mass there are also two 
solutions admitting a zero mode. One of them is the original RS solution with 
zero boundary masses, while the other, requiring nonvanishing boundary mass 
terms, can be termed the {\it specular} RS solution: its zero mode wave 
function, $\sim 1/A^2(z)$ and peaked on the IR brane, is the inverse of the 
original RS solution's wave function,  $A^2(z)$; see Sections 3-5 and Fig.1 
[specular: {\it (adj.) of, relating to, or having the qualities of a mirror}].}
The quadratic action of the graviton zero mode  for the  two branches is 
given in Appendix B. At the massive level, we also obtain the Kaluza-Klein 
spectrum. Most importantly, we note that the  scalar spectrum vanishes 
identically. Details of the quadratic action for scalar perturbations and 
the relevant boundary conditions are given in Appendix C.
\item We continue, in Section 4, with a discussion of the properties of the 
massless modes. We show that the tensor mode can be localized anywhere in 
the bulk and gives rise to a finite 4D Planck mass. This then motivates us 
to consider localizing the graviton on the IR brane and to interpret the 
force it mediates as an emergent infrared phenomenon since there is
no propagating massless spin 2 mode in the UV. In Sections 4.3, 4.4, and 
Appendix D, we study Newton's law at short distances and show that it  
exhibits a new power law dependence, that can even be irrational.
The crossover to a $V \propto 1/r^{\beta}$, $\beta >1$ (and real) behavior 
is observable at distances shorter than the infrared scale---but much larger 
than the AdS curvature scale, unlike the RS model---which can be taken to be 
in the sub-millimeter range. Gravity at such distances is presently under 
experimental investigation~\cite{gravexp1}-\cite{gravexp4}.

\item In  Section 5, we discuss the 4D holographic interpretation of our gravity solution.
 While it is necessarily speculative, even the remotest  plausibility  of a 4D dual picture 
inspires some confidence in the existence of a nonlinear extension of our linearized gravity analysis.
\begin{itemize}
\item In the first of the two branches---the one smoothly connected to the 
RS model---the dual  involves a strongly coupled 4D CFT coupled to a source 
graviton (Section 5.1). The coupling is relevant precisely when the zero mode 
is localized near the IR brane. As a result of this coupling, the energy 
momentum tensor of the CFT   should somehow acquire a large anomalous 
dimension. The conformal symmetry is broken at the infrared scale where a 
large Newton constant is induced by the CFT. The massless zero mode is thus 
a mixture of the source graviton and a CFT composite. The dominant component 
of the zero mode graviton, in the case of relevant coupling,  comes from the 
broken CFT. We show how the change in Newton's law can be explained in detail 
in the dual picture.
\item The holographic dual of the other branch of the gravity solution,  
described in Section 5.2, where the graviton is always localized near the 
IR brane, is even more mysterious. Holography suggests that it involves a  
CFT coupled to a massive source graviton (with mass of order the AdS 
curvature scale). Thus the observed massless graviton mode is essentially 
``emergent'' at low energies due to the strong infrared dynamics of the 
broken  CFT.
\end{itemize}
\item We conclude, in Section 6,  with a discussion of the effects of 
nonlinear terms to our analysis whose details are beyond the scope of the 
present paper.
\end{itemize}
Finally, let us comment on the relation of our model to  the ``fat"  graviton 
idea of Sundrum~\cite{fatgrav}, which was designed to turn  off gravity at 
short distances (precisely at the scales close to the experimental 
limit~\cite{gravexp1}-\cite{gravexp4}) and alleviate the cosmological 
constant problem.  In the picture presented here---which is the first 
quantitative model of a ``fat'' graviton, where gravity in many cases is 
an ``emergent'' low-energy phenomenon---the gravitational interaction  
becomes stronger at short length scales. The stronger UV gravity may be a 
more general phenomenon, or be simply due to the particular setup: gravity 
in a slice of AdS$_5$, being dual to a large-$N$ CFT, has a tower of stable 
``mesons" (the graviton Kaluza-Klein modes) contributing to the gravitational 
potential at short length scales, and may just signify that the 
``fat" graviton, without the effects of the Kaulza-Klein modes, 
should be looked for elsewhere. 

\section{Warped gravity with a bulk mass}

\subsection{The Randall-Sundrum background solution}

In the absence of mass terms,  warped gravity in five dimensions is governed
by the following 5D action:
\begin{equation}
   S = \int d^5 x \sqrt{-g} \left(M^3 R - 2\Lambda\right)
 -\sum_i \int d^4x\, \sqrt{-\gamma_i} \left( M^3 \left[ K \right] 
+ \lambda \right)_i ~,
\label{totalaction}
\end{equation}
where $M$ is the 5D Planck scale, $\Lambda$ is a bulk cosmological
constant, and the sum is over two boundary three-branes with brane 
tensions $\lambda_i$. The quantity $\left[ K \right]$ denotes the jump of the trace of the 
extrinsic curvature across the brane.

The solution to Einstein's equations is a slice of AdS$_5$ where the 
fifth dimension is compactified on an orbifold $S^1/Z_2$ of radius $R$ 
with the Randall-Sundrum metric~\cite{rs}:
\begin{equation}
   ds^2 = e^{-2 k y}\eta_{\mu\nu} dx^\mu dx^\nu + dy^2 
        = A^2(z)(\eta_{\mu\nu}dx^\mu dx^\nu+dz^2)
        \equiv g_{AB}^{(0)} dx^A dx^B~,
\end{equation}
where $0 \leq y\leq \pi R$ is the ``fundamental domain''
(where the bulk integral~(\ref{totalaction}) is computed),
$k$ is the AdS curvature scale, 
$\Lambda=-6k^2M^3$, and 
the Minkowski metric $\eta_{\mu\nu}$ has signature $(-+++)$. 
The Latin indices $(A,B,\dots)$ label all the 5D coordinates, while Greek 
indices $(\mu,\nu,\dots)$ are restricted to the 4D coordinates.
We will work with conformal coordinates defined by
$z=(e^{k y}-1)/k$ and $A(z) = ( 1 + k z)^{-1}$ is the warp factor.
At the orbifold fixed points $z_0=0$ and $z_1=(e^{\pi kR}-1)/ k$ 
there are two three-branes, the UV and the IR brane, respectively. 

In the 5D bulk the action is invariant under general coordinate 
transformations. At linear order the infinitesimal coordinate 
transformations are:
\begin{equation}
 x^M \rightarrow x^M + \xi^M(x)~,
\label{inf}
\end{equation}
where it is convenient to split the transformation parameters
into $\xi^M=(\xi^\mu+\partial^\mu\xi,\xi^5)$ with the 
condition $\partial_\mu \xi^\mu = 0$.
In this way the coordinate transformations are separated into two scalar 
$(\xi,\xi^5)$ and one vector $(\xi^\mu)$ transformation parameters 
(with respect to the Poincar{\'e} symmetry 
of the 4D Minkowski background).
In a covariant theory these can be used to eliminate five degrees of 
freedom (d.o.f).

The metric perturbations, $h_{AB}$ around the background Randall-Sundrum 
metric $g_{AB}^{(0)} \,$, correspond to fifteen degrees of freedom, and a 
useful way to parametrize them is:
\begin{eqnarray}
  ds^2 &=& A^2(z)\left[(1 + 2\phi) d z^2 + 2\,(B_{,\mu} + B_\mu) dz\,dx^\mu 
   \right.\nonumber\\
 &&\qquad\quad +\left.\left((1 + 2 \psi)\eta_{\mu \nu} +2E_{,\mu \nu} + 
   E_{\left( \mu,\nu\right)} + {\widehat h}_{\mu\nu} \right) 
    d x^\mu d x^\nu \right]~,\\
   &\equiv& (g_{AB}^{(0)} + h_{AB}) dx^A\, dx^B~,
\label{metdef}
\end{eqnarray}
where $E_{(\mu,\nu)}\equiv \partial_\mu E_\nu + \partial_\nu E_\mu$.
We see that the metric perturbations are divided into three sectors: 
scalar, vector, and tensor 
(with respect to the Poincar{\'e} symmetry 
of the 4D Minkowski background). 

Specifically, the tensor mode ${\widehat h}_{\mu\nu}$, is taken to satisfy 
the transverse ($\partial^\mu {\widehat h}_{\mu\nu}=0$) and traceless 
(${\widehat h}^\mu_\mu= 0$) conditions. It is gauge invariant under the 
infinitesimal coordinate transformations~(\ref{inf}), and being symmetric, 
it contains $10 - 5 = 5$ d.o.f. 

The vector modes $B_\mu$, and $E_\mu$ are both taken to be 
transverse ($\partial^\mu B_\mu = \partial^\mu E_\mu = 0$), and 
consequently contain $6$ d.o.f. Under the coordinate 
transformation~(\ref{inf}) the two vector modes transform as:
\begin{eqnarray}
&& B_\mu \rightarrow B_\mu - \xi_\mu'~,\\
&& E_\mu \rightarrow E_\mu - \xi_\mu~,
\label{infvec}
\end{eqnarray}
where prime ($'$) denotes differentiation with respect to $z$. Notice that
there is a gauge invariant combination:
\begin{equation}
   {\widehat B}_\mu \equiv B_\mu - E_\mu'~.
\label{vecgi}
\end{equation}
In a covariant theory, the orthogonal combination can be 
eliminated with 
$\xi_\mu$ using the coordinate transformation~(\ref{infvec}), leaving
only the three polarizations contained in ${\widehat B_\mu}$.

Lastly, the scalar modes $\psi,\phi, B$, and $E$, represent four real
degrees of freedom. 
Under the remaining infinitesimal coordinate transformations they 
transform as:
\begin{eqnarray}
&& \psi \rightarrow \psi - \frac{A'}{A} \, \xi^5~,\\
&& \phi \rightarrow \phi - \xi^{5'} - \frac{A'}{A} \, \xi^5~,\\
&& B \rightarrow B - \xi' - \xi^5~,\\
&& E \rightarrow E - \xi~.
\end{eqnarray}
There are now two gauge invariant combinations:
\begin{eqnarray}
&& {\widehat \psi} \equiv \psi - \frac{A'}{A} \left( B - E' \right)~,
\label{psihat} \\
&& {\widehat \phi} \equiv \phi - \frac{A'}{A} \left( B - E' \right) 
- \left( B - E' \right)'~. \label{phihat}
\end{eqnarray}
These two modes ${\widehat \psi}$, and ${\widehat \phi}$ represent 
two polarizations in the case of a covariant theory.

For later convenience, we also define the gauge invariant brane
positions. In general, a boundary brane can be at the perturbed position 
$z_i + \zeta_i \,$, and, as we shall see, the perturbation $\zeta_i$ couples 
to the scalar perturbations of the metric. Under the infinitesimal change 
of coordinates~(\ref{inf}), $\zeta_i \rightarrow \zeta_i + \xi^5 \,$. Hence, we can form 
the invariant combination:
\begin{equation}
{\widehat \zeta} \equiv \zeta + B - E' \,\,.
\label{zetahat}
\end{equation}

In summary, the perturbations of the metric contain 
$5 + 6 + 4 = 15$ degrees of freedom in the three sectors,
corresponding to the degrees of freedom of a symmetric $5 \times 5$ tensor. 
In the case of a covariant theory, such as the original Randall-Sundrum model,
they are reduced to $5 + 3 + 2 = 10$ polarizations. 
Not all of them are necessarily dynamical, since 
(as we will see) some polarizations vanish due to constraint equations 
that result from the equations of motion (equivalently, they can be seen 
as Lagrange multipliers in the original action). This parametrization of the
perturbations will be also useful when we add mass terms to the action.

\subsection{Adding a bulk mass term}

Let us now consider adding a mass term for the perturbations 
$h_{A B} \equiv g_{A B} - g_{A B}^{(0)} \,$, where $g_{AB}^{(0)}$ is the 
background Randall-Sundrum metric. The 5D bulk action
becomes:
\begin{equation}
 S = \int d^5 x \sqrt{-g} \left[ M^3 R - 2\Lambda -M^3\, k^2 \, 
    g^{(0)\,M N} \, g^{(0)\,A B} \left( a \, h_{M A} \, h_{N B} 
    + b \, h_{M N} \, h_{A B} \right) \right]~,
\label{buaz}
\end{equation}
where $a$ and $b$ are real parameters. The previous classification of the 
perturbations $h_{AB}$ is useful since modes belonging to different 
representations are not coupled to each other at the linearized level. This means
that we can write the Einstein equations in the three sectors independently. 
For the covariant case, it is easiest to compute the equations in the form
$\delta G^M_N = 0$, where $\delta G$ is the linear perturbation of the 
Einstein tensor. This equation holds as long as there is only a 
cosmological constant in the bulk (adding dynamical fields in the bulk gives 
rise to a nonvanishing $\delta T^M_N \,$).

If we now include the bulk mass term, then the action (\ref{buaz}) 
leads to the bulk Einstein equations:
\begin{equation}
\delta G^A_B + 2 M^3\, k^2 \left[ a \, h^A_B + b \, h \, \delta^A_B \right] 
= 0~,
\label{bulk}
\end{equation}
where $h^A_B = g^{(0) A C} \, h_{C B} \,$ and $h = h^A_A \,$. Let us now
separately consider the nontrivial Einstein equations in each of the three 
sectors:

\subsubsection{Tensor}

The equation of motion for the tensor modes ${\widehat h}_{\mu\nu}$ is 
the transverse--traceless part of the $\mu \nu$ component of~(\ref{bulk}),
and is given by:
\begin{equation}
    \Box {\widehat h}_{\mu\nu} + {\widehat h}_{\mu\nu}''
      + 3 \, \frac{A'}{A} \, {\widehat h}_{\mu\nu}' 
      - 4 \,a \, k^2 A^2 \, {\widehat h}_{\mu\nu}= 0~,
\label{butmn}
\end{equation}
where $\Box \equiv - \partial_t^2 + \partial_{\bf x}^2$. Notice that 
this equation does not depend on the $b$ part of the mass term (\ref{buaz})
since the tensor mode ${\widehat h}_{\mu\nu}$ is traceless.
The solution of the equation of motion is obtained by a separation of 
variables ${\widehat h}_{\mu\nu}(x,z)= f(z) H_{\mu\nu}(x)$, where 
$\Box H_{\mu\nu}(x) = m^2 H_{\mu\nu}(x)$, with $m$ representing the mass 
of the four-dimensional Kaluza-Klein modes. The massless mode solution is:
\begin{equation}
   {\widehat h}_{\mu \nu}^{(0)}(x,z) = \left[ C_1 \, A(z)^{-2 \left( 1 
    - \sqrt{1+a} \right)} + C_2 \, A(z)^{-2 \left( 1 
    + \sqrt{1+a} \right)} \right] H_{\mu \nu}^{(0)}(x)~,
\label{tbulknomass}
\end{equation}
while the massive modes are:
\begin{equation}
    {\widehat h}_{\mu \nu}^{(n)}(x,z) = A^{-2}(z)
     \left[  \, C_1 \, J_{2 \, \sqrt{1+a}} \left( \frac{m_n}{k A(z)} 
     \right) +  C_2 \, Y_{2 \, \sqrt{1+a}} 
     \left( \frac{m_n}{k A(z)} \right)\right] \, 
     H_{\mu \nu}^{(n)}(x)~,
\label{massten}
\end{equation}
where $C_1,C_2$ are arbitrary constants. We will consider only values
$a \geq -1 \,$, which include the Randall-Sundrum case ($a=0$), and, as we 
will see, provides a general and interesting phenomenology.
In the limiting case $(a=-1)$, the massless solutions are degenerate.
Also note that in the limit $a\rightarrow 0$ these modes become:
\begin{eqnarray}
   {\widehat h}_{\mu \nu}^{RS,(0)}(x,z) &=& \left[ C_1 
    + C_2 \, A(z)^{-4} \right] H_{\mu \nu}^{(0)}(x)~, \label{modet0} \\
    {\widehat h}_{\mu \nu}^{RS, (n)}(x,z) &=& A^{-2}(z)
     \left[  \, C_1 \, J_2 \left( \frac{m_n}{k A(z)} 
     \right) +  C_2 \, Y_2 \left( \frac{m_n}{k A(z)} \right)\right] \, 
     H_{\mu \nu}^{(n)}(x)~,
     \label{modetm}
\end{eqnarray}
which, together with appropriate boundary conditions (see below)
smoothly reproduce the Randall-Sundrum solution~\cite{rs}.

\subsubsection{Vector}

The $5 \mu$ and $\mu \nu$ components of~(\ref{bulk}) lead to the Einstein
equations for the vector modes:
\begin{eqnarray}
 \Box {\widehat B}_\mu - 4 \,a \, k^2 A^2\, B_\mu &=& 0~, \label{buv5m} \\
 {\widehat B}_\mu^{'} + 3 \, \frac{A'}{A} \, {\widehat B}_\mu + 4 \,a \, k^2 
A^2 \, E_\mu &=& 0~.
\label{buvmn}
\end{eqnarray}
These equations can be decoupled by eliminating $B_\mu$ using (\ref{vecgi})
to obtain equations which depend only on ${\widehat B}_\mu$ and $E_\mu$.  
This gives rise to a second order equation solely in terms of 
${\widehat B}_\mu$:
\begin{equation}
  \Box{\widehat B}_\mu+{\widehat B}_\mu'' - k A \, {\widehat B_\mu}' 
   - (4\, a +3)\, k^2 A^2 {\widehat B}_\mu = 0~,
\label{Bhateqn}
\end{equation}
where (\ref{buvmn}) can be used to obtain $E_\mu\,$ from ${\widehat B}_\mu$.
As for the tensor mode we see that the vector modes do not depend on 
the $b$ part of the mass term (\ref{buaz}) since the vector modes are
transverse and do not contribute to the trace of the perturbation
in the mass term.
The equation (\ref{Bhateqn}) is again solved by separating the variables.
When $a\neq 0$ the massless mode solutions are:
\begin{eqnarray}
{\widehat B}_\mu^{(0)}(x,z) &=& \left[ C_1 A(z)^{-1-2 \sqrt{1+a})} 
   + C_2 A(z)^{-1+2 \sqrt{1+a}} \right] b_\mu^{(0)}(x)~,
\label{vbulknomass}\\
E_\mu^{(0)}(x,z) &=& -\frac{1}{2\,a\,k}\left[ C_1 \left( 1 - \sqrt{1+a} 
\right) A(z)^{-2 \left( 1 + \sqrt{1+a} \right)}\right.\nonumber\\
&&\qquad\qquad \left. + \, C_2
\left( 1 + \sqrt{1+a} \right) A(z)^{-2 \left( 1 - \sqrt{1+a} \right)} 
\right] b_\mu^{(0)}(x)~,
\end{eqnarray}
while the massive mode solutions are given by:
\begin{equation}
{\widehat B}_\mu^{(n)}(x,z) = A^{-1}(z) \left[ C_1 \, J_{2 \, \sqrt{1+a}} 
    \left( \frac{m_n}{k A(z)} \right) + C_2 \, Y_{2 \, \sqrt{1+a}} 
    \left( \frac{m_n}{k A(z)} \right) \right] \,b_\mu^{(n)}(x)~,
\label{massivevec}
\end{equation}
where $C_1,C_2$ are arbitrary constants and $E_\mu^{(n)}$ are obtained 
from (\ref{buvmn}). 

We can compare this with the massless case. When $a=0$ the equations of motion 
(\ref{buv5m}) and (\ref{buvmn}) only depend on ${\widehat B}_\mu$. 
From (\ref{buv5m}) we see that the vector mode is always 
massless and since (\ref{buvmn}) is a first order differential equation there
is only one solution given by:
\begin{equation}
{\widehat B}_\mu^{RS}(x,z) = C_1 A^{-3}(z)\, b_\mu(x)~,
\label{RSvec}
\end{equation}
where $C_1$ is an arbitrary constant. We will see later that the 
boundary conditions will eliminate this mode, as is well known for the 
RS case.

\subsubsection{Scalar}

The scalar equations are obtained from the $55 \,$, $5 \mu$ and $\mu \nu$ 
components of~(\ref{bulk}). This leads to the following equations involving
the scalar modes:
\begin{eqnarray}
 \Box {\widehat \psi} + 4 \, \frac{A'}{A} \left( {\widehat \psi}' - 
\frac{A'}{A} \, {\widehat \phi} \right) + \frac{4}{3}\, k^2 A^2\left[ a \, 
\phi + b \left( 4\, \psi + \, \phi + \Box E \right) \right] &=& 0~, 
\label{bus55} \\
 {\widehat \psi}' - \frac{A'}{A} \, {\widehat \phi} 
- \frac{2}{3} \, a \, k^2 A^2\, B &=& 0~, 
\label{bus5m}\\
{\widehat \phi} + 2 \, {\widehat \psi} - 4\,a\,k^2 A^2\,E &=& 0~,
\label{busmn1}\\
\frac{1}{3}\Box \left( {\widehat \phi} + 2 \,{\widehat \psi} \right)
+ {\widehat \psi}'' + 3 \, \frac{A'}{A} \, {\widehat \psi}' 
- \frac{A'}{A} \, {\widehat \phi}' - 4 \frac{A^{'2}}{A^2} \, {\widehat \phi} 
\qquad\qquad&& \nonumber\\
+ \frac{4}{3} \, k^2 A^2\left[ a \, \psi + b \left( 4 \, \psi + \phi 
+ \Box E \right) \right] &=& 0~.
\label{busmn2}
\end{eqnarray}
Notice that unlike the vector and tensor modes the coefficient $b$ now
appears in the scalar equations. This is because the scalar modes do
contribute to the trace of the metric perturbations. 

When $a=b=0$, only the
gauge invariant combinations of the metric perturbations appear in the 
Einstein equations. The system of equations is straightforward to
solve and from (\ref{bus5m}) and (\ref{busmn1}) we obtain:
\begin{equation}
   {\widehat \phi} = - \, 2 \, {\widehat \psi}~,\qquad\qquad
{\widehat \psi}(x,z) = C_1 \, A^{-2}(z) \, S(x)~,
\label{radion}
\end{equation}
where $C_1$ is an arbitrary constant.
Hence, as in the vector sector, only one mode is present in the 
RS case. Using equation (\ref{bus55}) this mode (the radion)
is massless, $\Box S(x) = 0$.
The remaining equation (\ref{busmn2}) is degenerate and consistent with
the solution (\ref{radion}).

The solution of the system of equations (\ref{bus55})-(\ref{busmn2})
for the scalar perturbations $\psi,\phi, B$ and $E$ when $a\neq 0$ and 
$b\neq 0$ is more involved. As first step we will eliminate $\psi$ and 
$\phi$ in terms of ${\widehat \psi}$ and ${\widehat \phi}$. Thus, using 
(\ref{psihat}), ({\ref{phihat}), (\ref{bus5m}), and 
(\ref{busmn1}) we obtain:
\begin{eqnarray}
E &=&  \frac{1}{4 \, a \, k^2 A^2} \left( {\widehat \phi} + 2 \, 
   {\widehat \psi} \right)~,
\label{eeqn}\\
B &=&  \frac{3}{2 \, a \, k^2 A^2} \left( {\widehat \psi}' 
   + k A \, {\widehat \phi} \right)~, \\
\psi &=& {\widehat \psi} - \frac{1}{4 \, a \, k A} \left[ 4\,{\widehat \psi}' 
   - {\widehat \phi}' + 4 \, k A \left( {\widehat \phi} - {\widehat \psi} 
   \right) \right]~, \\
\phi &=& {\widehat \phi} + \frac{1}{4 \, a \, k^2 A^2} \, \left( 4 \, 
    {\widehat \psi}'' - {\widehat \phi}'' + 3 \, k A {\widehat \phi}' 
\right)~.
\label{phieqn}
\end{eqnarray}
Using these equations the remaining two scalar equations, (\ref{bus55}) 
and (\ref{busmn2}) can then be written in terms ${\widehat \psi}$ and 
${\widehat \phi}$. This leads to two coupled second order differential 
equations which can be solved for ${\widehat \psi}$ and ${\widehat \phi}$. 
These equations are equivalent to a fourth order differential equation, 
which is difficult to solve in general. However, the coupled 
differential equations magically simplify for the ``Fierz--Pauli'' 
choice:\footnote{In fact the major motivation for this choice is 
provided by the presence of $\left( \Box E \right)^2$ in the quadratic 
action for the scalar perturbations. As in the 4D case~\cite{ags}, these 
terms cancel for the Fierz--Pauli choice; see Appendix C.}
\begin{equation}
 b=-a~, 
\label{pf}
\end{equation}
of the bulk mass parameters in~(\ref{buaz}), since the sum of the two coupled
equations only involves the combination ${\widehat \psi} - {\widehat \phi}$.
The orthogonal combination ${\widehat \psi} + {\widehat \phi} \,$ is then 
most easily obtained from~(\ref{bus55}). Therefore, defining the linear
combinations:
\begin{equation}
X \equiv {\widehat \psi} - {\widehat \phi} \;\;\;,\;\;\; 
Y \equiv {\widehat \psi} + {\widehat \phi} \,\,,
\label{xiy}
\end{equation}
the remaining two scalar equations are:
\begin{eqnarray}
&& \Box \, X + X'' +  k A \, X' - 4 \, (1 + a)\,k^2 A^2 \, X=0~,
\label{xieqn}\\
&& Y=\frac{1}{\left( 3 +4 \, a \right)\,k^2 A^2} \, \left[
2\, k A \, X' - ( 5+4\,a)\, k^2 A^2 \,X + \frac{1}{2}\Box\,X\right]~.
\label{Yeqn}
\end{eqnarray}
Hence, we have obtained a second order equation in terms of $X$ only.
The remaining equation is an algebraic equation for $Y$, which is trivially
solved in terms of $X$. It is straightforward to obtain the general solution of 
the remaining differential equation (\ref{xieqn}).
Again separating the variables and writing $X(x,z)=f(z)S(x)$ we find that
the massless modes solving these equations are:
\begin{eqnarray}
X^{(0)}(x,z) &=& \left[ C_1 A(z)^{-2 \sqrt{1+a}} 
+ C_2 A(z)^{2 \sqrt{1+a}} \right] S^{(0)}(x)~,
\label{sbulkxi}\\
Y^{(0)}(x,z) &=&\frac{-1}{4a+3} \left[ C_1 \left(2 \sqrt{1+a} 
-1 \right)^2 A(z)^{-2\sqrt{1+a}} \right. \nonumber\\
&&\qquad\qquad \left. +\,C_2 \left(2 \sqrt{1+a} + 1 \right)^2 
A(z)^{2\sqrt{1+a}} \right] S^{(0)}(x)~,
\label{sbulkY}
\end{eqnarray}
where $C_1,C_2$ are arbitrary constants and 
the four dimensional mode obeys $\Box S^{(0)}(x) = 0$. If we now
substitute these general 
solutions back into (\ref{eeqn})-(\ref{phieqn}) we obtain:
\begin{eqnarray}
E^{(0)}(x,z) &=& \frac{-1}{2\,a\,\left(3 + 4\,a \right) \,k^2}
    \left[ C_1 (2\,a + 3(1 - \sqrt{1+a}))\, A(z)^{-2(1-\sqrt{1+a})}\right.
\nonumber\\
&&\qquad \left.
+\,C_2 (2\,a + 3\,( 1 + \sqrt{1+a})) \, A(z)^{-2(1+\sqrt{1+a})}\right]
S^{(0)}(x)~,
\label{E0gensoln}\\
B^{(0)}(x,z)&=&0~,\\
\psi^{(0)}(x,z)&=&0~,\\
\phi^{(0)}(x,z)&=&0~.
\end{eqnarray}
Remarkably, the bulk equations of motion have forced all the massless scalar
perturbations to become zero, except for the $E^{(0)}$ mode! 
Furthermore, as shown in Appendix~\ref{appC}, this mode gives a vanishing 
contribution to the action for the perturbations, and therefore 
is not physical (at least, to quadratic order in the perturbations). 
Thus, the addition of the bulk mass term with the Fierz--Pauli 
choice (\ref{pf}) has completely eliminated the massless (radion) mode.

The massive mode solutions are:
\begin{equation}
X^{(n)}(x,z) = \left[ C_1 \, J_{2 \, \sqrt{1+a}} 
    \left( \frac{m_n}{k A(z)} \right) + C_2 \, Y_{2 \, \sqrt{1+a}} 
    \left( \frac{m_n}{k A(z)} \right) \right] \,S^{(n)}(x)~,
\label{massxin}\\
\end{equation}
where $\Box \, S^{(n)}(x)= m^2 S^{(n)}(x)$ and $Y^{(n)}(x,z)$ is obtained from
(\ref{Yeqn}).
As we shall see, also these modes disappear, once the boundary conditions are
taken into account.

\section{Brane localized mass terms}

\subsection{Boundary conditions}

In order to determine the mass spectrum from the general solutions we need to 
derive the boundary conditions satisfied by the bulk modes on the branes.
We will also add boundary mass terms on the branes since this will be crucial
for obtaining a deformation of the RS solution. Hence, consider
the following brane action at the location $z_i$:
\begin{equation}
\Delta S_i = -k \,M^3\, \int d^4 x \, \sqrt{- \, \gamma_0} \, h_{\mu \nu} \, 
h_{\alpha \beta} \left( \alpha_i \, \gamma_0^{\mu \alpha} \, 
\gamma_0^{\nu \beta} 
+ \beta_i \, \gamma_0^{\mu \nu} \, \gamma_0^{\alpha \beta} \right)~,
\label{fpbound}
\end{equation}
where $\gamma_{0,\mu \nu} = A^2 \, \eta_{\mu \nu}$ is the background induced 
metric on the boundary, with $A$ evaluated at the (unperturbed) location of 
the brane, and $h_{\mu \nu}$ are the perturbations of the induced metric,
\begin{equation}
\gamma_{\mu \nu} = \gamma_{0,\mu \nu} + h_{\mu \nu} \,\,.
\end{equation}
The total induced metric $\gamma$ is evaluated at the perturbed brane
position $z_i+\zeta_i \left( x^\mu \right) \,$. The 
brane displacements $\zeta_i$ constitute two additional scalar modes of 
the system which have support only on the two boundaries. It is worth noting 
that the action~(\ref{fpbound}) is not the unique boundary action which one 
could choose. This term itself is not covariant, and there are several 
inequivalent possibilities (as we discuss below in Section~\ref{subsca}).
Our choice was motivated by the fact that 
(\ref{fpbound}) is rather simple and natural.

The boundary mass terms give a contribution to the energy-momentum tensor 
on the boundary:
\begin{equation}
\delta S_{\mu \nu} \equiv - \, \frac{2}{\sqrt{- \, \gamma_0}} \, 
\frac{\delta \, \Delta S_i}{\delta h^{\mu \nu}} = - \, 4 \, k \, M^3
\left( \alpha \, h_{\mu \nu} + \beta \, h \, \gamma_{0\; \mu \nu} \right)~,
\label{em1}
\end{equation}
where to linear order in the perturbations:
\begin{equation}
   h_{\mu\nu} = 2 A^2(z_i)\left[\left(\psi
  +\frac{A'(z_i)}{A(z_i)}\zeta_i\right)\eta_{\mu\nu} + E_{,\mu\nu} + 
\frac{1}{2} \left( E_{\mu , \nu} + \widehat{h}_{\mu\nu}\right)\right]~,
\label{hus}
\end{equation}
and $h = h^\mu_\mu = \gamma_0^{\mu \nu} \, h_{\nu \mu} \,$, with
$\gamma_0^{\mu \nu}$ the inverse background induced metric.
This contribution must be added to the standard RS piece
arising from the brane tensions, namely $S_{\mu \nu}^{(0)}=-\lambda_i 
\gamma_{\mu\nu}$, where $\lambda_i=\pm\, 6\,k\,M^3\,$ and $\pm$ refers to the 
UV/IR brane, respectively. Hence the total energy--momentum tensor is:
\begin{equation}
 S_{\mu \nu} = S_{\mu\nu}^{(0)} + \delta S_{\mu\nu} \,\,.
\end{equation}
This expression is related to the jump of the extrinsic curvature 
$K_{\mu \nu}$ across each brane. Note that the extrinsic curvature is 
computed from bulk quantities, and formally it is not affected by the 
bulk mass term in~(\ref{buaz}). 
The extrinsic curvature, up to first order in the perturbations, is given by: 
\begin{equation}
K_{\mu \nu} = \nabla_\mu \, n_\nu = \partial_\mu n_\nu - \, \Gamma^5_{\mu \nu} \, n_5 
   = \partial_\mu n_\nu- \, \Gamma^5_{\mu \nu} \, A \, 
   \left( 1 + \phi +\frac{A'}{A}\zeta_i\right) \,\,.
\end{equation}
Evaluating it from the bulk geometry gives:
\begin{eqnarray}
K_{\mu \nu} &=& A' \, \eta_{\mu \nu} + \delta K_{\mu \nu} \,\,, \nonumber\\
\delta K_{\mu \nu} 
&=& A \Bigg\{ \left( \psi' + \frac{2 \, A'}{A} \, \psi 
- \frac{A'}{A} \, \phi +\frac{A''}{A}\zeta_i\right) \eta_{\mu \nu}
+ \left( E' + \frac{2 A'}{A} \, E - B -\zeta_i\right)_{,\mu \nu} \nonumber\\
&& \quad + \left( \frac{1}{2} E'_{(\mu} + \frac{A'}{A} E_{(\mu} 
- \frac{1}{2} B_{(\mu} \right) \mbox{}_{,\nu)} 
+ \left( \frac{1}{2} {\widehat h}'_{\mu\nu} + \frac{A'}{A} 
{\widehat h}_{\mu\nu}
\right) \Bigg\}~.
\end{eqnarray}
The junction conditions are then:
\begin{equation}
M^3 \left[ {\widehat K}_{\mu \nu} \right] = - S_{\mu \nu}~,
\label{jump}
\end{equation}
where ${\widehat K}_{\mu\nu}=K_{\mu\nu}-K h_{\mu\nu}$ and 
$\left[ \dots \right]$ means the jump across the brane (with $Z_2$ symmetry 
imposed). Evaluating the junction conditions explicitly in the 
three sectors gives:
\begin{eqnarray}
\!\!\!\!\!\! {\rm tensor:} && \!\! {\widehat h}'_{\mu\nu} = 
\pm 4 \,\alpha_i \, k \, A \, {\widehat h}_{\mu\nu}~, \label{junt} \\
\!\!\!\!\!\! {\rm vector:} && \!\! {\widehat B}_\mu = \mp
4 \,\alpha_i \, k \, A \, E_\mu~, \label{junv}\\
\!\!\!\!\!\! {\rm scalar:} && \psi' - \frac{A'}{A}\phi = 
\pm 4 k A \left[\alpha_i \left(\psi+\frac{A'}{A}\zeta_i\right) 
- \frac{\alpha_i + \beta_i}{3}\left(4\left(\psi+\frac{A'}{A}\zeta_i\right) 
+ \Box E \right) \right]~~ \label{juns1}\\
\!\!\!\!\!\! {\rm scalar:} && \!\! E' - B -\zeta_i = \pm
4 \,\alpha_i \, k \, A \, E~, 
\label{juns2}
\end{eqnarray}
These expressions are to be evaluated at the brane locations $z_0$ (upper sign) and $z_1$ (lower sign).
They are valid in general, both for the massless and massive modes, and we
will use them to determine the mass spectrum.

In the scalar sector, there is an additional boundary condition, which is 
obtained directly from varying the quadratic action for the scalar 
perturbations with respect to the brane displacement $\zeta_i \,$ 
(see Appendix~\ref{appC} for details). When the mass terms are absent, this 
equation is actually redundant with respect to the boundary conditions 
given above. This is not surprising since it is a consequence of 
general covariance of the model. In the present case, the inclusion of 
the mass terms removes this degeneracy, leading to the additional 
boundary condition. For $b = - a \neq 0 \,$, and $\beta_i = - \alpha_i 
\neq 0 \,$, the additional boundary condition can be rewritten as:
\begin{eqnarray}
\!\!\!\!\!\! {\rm scalar:} && \!\! 4 \, \psi + \Box E = 0 \,\,,
\label{juns3}
\end{eqnarray}
which, like the previous boundary conditions, only holds at the two boundaries.

\subsection{Tensor modes}

Applying the boundary condition (\ref{junt}) to the tensor
mode general solution~(\ref{tbulknomass}) gives:
\begin{equation}
(1 - \sqrt{1+a} \mp 2\alpha_i) C_1 \, A(z_i)^{2\sqrt{1+a}} +
(1 +  \sqrt{1+a} \mp 2\alpha_i) C_2 \, A(z_i)^{-2\sqrt{1+a}} = 0~,
\label{jt2}
\end{equation}
where this equation is evaluated at $z_0$ (upper sign) or $z_1$ (lower sign). For generic 
mass parameters $a,\alpha_i$ the only solution is $C_1=C_2=0$, and there
is no massless graviton. However, when:\footnote{Since we restrict ourselves to the Fierz--Pauli
choice $\beta_i = - \alpha_i \,$, we also set $\beta_0 = - \beta_1 \equiv \beta$ for the scalar modes.}
\begin{equation}
\alpha_0 = - \alpha_1 \equiv \alpha \;\;,\;\; \alpha = \frac{1}{2}(1\mp\sqrt{1+a})\equiv \alpha_\mp~,
\label{alpha}
\end{equation}
the coefficient $C_1\,(C_2)$ drops from the boundary condition, and (\ref{jt2})
simply gives $C_2 = 0\,(C_1=0)$. Hence, the massless tensor mode 
(\ref{tbulknomass}) becomes:
\begin{equation}
{\widehat h}_{\mu\nu} (x,z)=  N_T A(z)^{-4\alpha} \, 
H_{\mu \nu}^{(0)}(x)~,
\label{t0mode}
\end{equation}
where $N_T$ is the overall normalization constant which is determined from the 
quadratic action in the perturbations and cannot be determined by the 
boundary conditions. 

\begin{figure}
\centerline{
\includegraphics[width=0.65\textwidth]{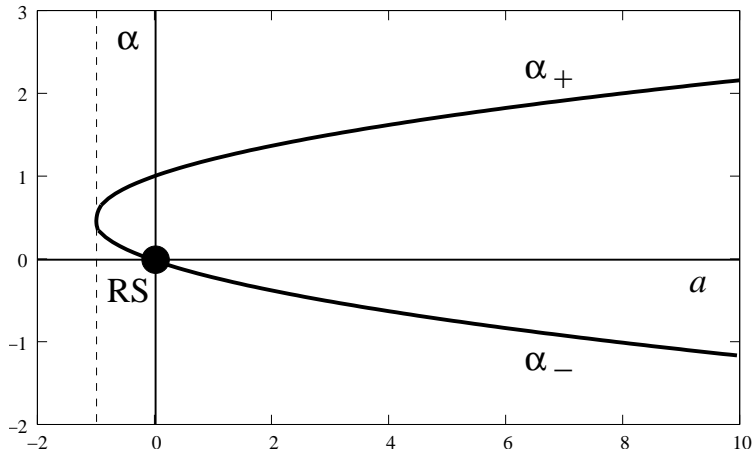} }
\caption{The range of bulk ($a$) and brane ($\alpha$) mass 
parameters that lead to a 4D massless graviton. There are two 
branches, $\alpha_\pm$, joined together at the limiting value $a=-1$. 
The RS model $(\alpha=a=0)$ is a special case on the 
$\alpha_-$ branch. Its mirror image on the $\alpha_+$ branch  is the ``specular" RS solution ($a=0$, $\alpha = 1$).}
\label{fig1}
\end{figure}

This form of the solution is only meaningful when
the bulk mass parameter $a\geq -1$. This corresponds to
$\alpha=\alpha_+ (\alpha_-)$ for $\alpha \geq 1/2 \, 
(\alpha \leq 1/2)$, so that the full range of the boundary mass parameter
$\alpha$ is covered by the two branches $\alpha_\pm$.
This behavior is plotted in Figure~\ref{fig1}.
When $a,\alpha_-\rightarrow 0$, the bulk and boundary
mass terms become zero, and the $\alpha_-$ mode reduces to the usual 
RS tensor mode that is constant in the $z$ coordinate.
Hence the $\alpha_-$ mode is a smooth deformation of the RS
tensor mode from $\alpha =0$ to $-\infty<\alpha \leq 1/2$. On the other hand 
the $\alpha_+$ mode can only exist when the boundary mass is nonzero 
and corresponds to the deformation of the RS 
tensor mode to values $1/2\leq \alpha <\infty$. Later, we will see 
that this corresponds to localizing modes continuously from the UV 
brane to the IR brane.

It is worth highlighting the presence of the massless zero mode for $a=0 \,$,
$\alpha_+ = 1 \,$, arising for a special value of the brane mass parameter, 
and without a mass term in the bulk. In this case, general covariance is 
broken only at the location of the two branes, while the 5D bulk is 
general coordinate invariant. As is clear in Fig.~\ref{fig1}, this model 
is the ``specular'' version of the RS mode. Indeed, as we have 
seen, the bulk equation~(\ref{butmn}) is a second order differential equation,
and in particular the value $a=0$ leads to 
two linearly independent bulk solutions. The choice of $\alpha_+ = 1$ leads 
to the survival of the massless mode that is normally removed by the 
RS boundary conditions (and vice versa).

For the massive solutions (\ref{massten}) the boundary condition (\ref{junt})
for $\alpha=\alpha_\pm$ can be written as
\begin{equation}
      \frac{J_{2\sqrt{1+a}\pm 1}
     \left(\frac{m_n}{k A(z_0)}\right)}
     {Y_{2\sqrt{1+a}\pm 1}\left(\frac{m_n}{k A(z_0)}\right)}
=\frac{J_{2\sqrt{1+a}\pm 1}
     \left(\frac{m_n}{k A(z_1)}\right)}
     {Y_{2\sqrt{1+a}\pm 1}\left(\frac{m_n}{k A(z_1)}\right)}
\label{masstenbc}~.
\end{equation}
Since $A(z_0)=1$ and $A(z_1)=e^{-\pi k R}$ we obtain in the limit 
$k\, e^{-\pi k R} \ll m_n \ll k$ the Kaluza-Klein mass spectrum
\begin{equation}
      m_n \simeq \left(n+\sqrt{1+a}-\frac{3}{4}\right)\, \pi\,k e^{-\pi k R}~,
\qquad n=1,2,3,\dots
\label{tenKKmass}~.
\end{equation}
This approximation for the mass spectrum becomes increasingly better 
as $n$ grows. When $a=0$ we recover the RS Kaluza-Klein mass 
spectrum~\cite{rs}.

\subsection{Vector modes}

Since the massless tensor mode only exists for $\alpha=\alpha_\pm$ we will
also assume this value of $\alpha$ for the vector mode.
The boundary condition (\ref{junv}) for the massless
vector mode (\ref{vbulknomass}) becomes:
\begin{equation}
 C_1 (1+\frac{4}{a}\alpha\,\alpha_-)\, A(z_i)^{-(1 + 2 \sqrt{1+a})} + 
   C_2 (1+\frac{4}{a}\alpha\,\alpha_+)\, A(z_i)^{-(1 - 2 \sqrt{1+a})}= 0~,
\label{vecbc}
\end{equation}
where this condition is imposed at $z_0$ and $z_1$. When $\alpha = \alpha_-$, 
the term proportional to $C_2$ in (\ref{vecbc}) 
vanishes (since $\alpha_+\alpha_- = -a/4$),
which then simply enforces $C_1 = 0 \,$. On the contrary, for 
$\alpha = \alpha_+$ the boundary condition (\ref{vecbc}) leads to $C_2=0$. 
Thus the massless vector mode solution becomes:
\begin{eqnarray}
  {\widehat B}_\mu^{(0)}(x,z) &=& N_V A(z)^{1-4\alpha} \,b_\mu^{(0)}(x)~,
\label{newvec}\\
  E_\mu^{(0)}(x,z)&=& \frac{N_V}{a\,k}~(1-\alpha)~A(z)^{-4\alpha} \,
b_\mu^{(0)}(x)~,
\end{eqnarray}
where $N_V$ is a normalization constant and $\alpha=\alpha_\pm$ for
$\alpha \geq 1/2\, (\alpha \leq 1/2)$. Notice that when 
$a,\alpha_-\rightarrow 0$
the mode which gets killed, $C_1=0 \,$, is precisely the solution which would 
have smoothly gone to the RS vector mode (\ref{RSvec}).
Hence, for $\alpha\leq 1/2$ (including, in particular, the RS 
limit $\alpha=0$) this mode is always killed by the $Z_2$ symmetry. 
However, we see that for nonzero boundary masses $\alpha\neq 0$ there 
is a new massless mode present, given by  (\ref{newvec}), that is instead absent in the
RS case. Just like the tensor mode, this mode
can be localized anywhere in the bulk.

The boundary condition (\ref{junv}) for the massive solutions 
(\ref{massivevec}) leads to the same equation (\ref{masstenbc}) as for
the graviton tensor modes, and hence to an identical Kaluza-Klein mass spectrum 
(\ref{tenKKmass}). This is not too surprising since both modes
originate from the five-dimensional graviton.

\subsection{Scalar modes}
\label{subsca}

Let us discuss the zero and massive scalar modes separately. For the zero 
modes, we have seen that the bulk equations enforce $\psi = \phi = B = 0 \,$. 
The boundary condition~(\ref{juns1}) then leads to $\zeta_i = 0$ for both 
branes. This leaves only the scalar mode $E$ which can in principle be 
nonvanishing. However, an explicit calculation shows that if $E$ is the 
only mode present, it does not contribute to the quadratic action for the 
perturbations. Hence we conclude that, at least at the linearized level, 
there are no massless scalar modes in the theory.\footnote{We 
have verified that for the ``specular" solution, $a=0$, $\alpha = 1$, 
where the calculation is slightly different, there is also no scalar mode 
in the spectrum.}

For the massive modes, there are three nontrivial boundary 
conditions~(\ref{juns1}), (\ref{juns2}), and (\ref{juns3}). Since each 
equation holds on both branes, we have a system of six 
equations in four variables for each massive mode. The variables are the 
displacements of the two branes $\zeta_{0,1}$ and from the 
mode~(\ref{massxin}), the ratio $C_1 / C_2$ and the mass $m_n$. 
Since this is an overdetermined system of equations we can only 
hope to have nontrivial massive scalar modes if some of these equations 
are degenerate.

To check if there is a degeneracy first note that, after some algebra, 
Eq.~(\ref{juns3}) forces the mode $X^{\left( n \right)}$ in 
(\ref{massxin}) to vanish at both branes. The remaining equations can then 
be combined to eliminate $\zeta_i$, and together with~(\ref{juns3}), they 
also force $X'$ to vanish at the two branes. Thus there is no degeneracy
and the requirements of:
\begin{equation}
X^{\left( n \right)} = X^{\left( n \right)'} = 0~,  \quad\quad {\rm at} \; 
{\rm both} \; {\rm boundaries}~,
\end{equation}
can only be satisfied for $C_1 = C_2 = 0 \,$. This immediately implies that 
the massive scalar modes are also absent.

The absence of massive scalar modes is due to a ``mismatch'' between the 
bulk and boundary mass terms that we have introduced. Both terms can be 
considered as massive deformations of the original Randall--Sundrum 
proposal. We have previously shown that the deformation~(\ref{em1}) 
of the boundary action ``matches'' with the deformation~(\ref{buaz}) 
of the bulk action to produce nontrivial tensor and vector modes. However,
we now see that this is not the case for the scalar sector. It is possible 
that some inequivalent choice for the boundary actions can also accommodate 
nonvanishing scalar modes. As we already mentioned, the action~(\ref{em1}) 
is not covariant, so other inequivalent actions can be considered. This 
is particularly relevant for the scalar modes, which have the additional 
ambiguity (with respect to the other sectors) of the brane positions. For 
instance, the perturbation of the induced metric in~(\ref{em1}) is
defined as:
\begin{equation}
h_{\mu \nu} = \gamma_{\mu \nu} - A^2 \left( z_i \right)\, \eta_{\mu \nu}~,
\label{hus2}
\end{equation}
where $\gamma_{\mu \nu}$ is the (total) induced metric, with 
scalar perturbations included. In this way, the background induced metric 
is evaluated at the unperturbed brane positions $z_{0,1}$. At linear order, 
this leads to the expression~(\ref{hus}). Alternatively, one can choose 
to evaluate also the second term in~(\ref{hus2}) at the perturbed brane 
positions $z_i + \zeta_i$. This results in omitting the terms proportional 
to $\zeta_i$ in the expression~(\ref{hus}). As a consequence, eq.~(\ref{juns1})
also appears without the two terms proportional to $\zeta_i$, while clearly the
vector and tensor sectors are instead unaffected by this choice. We have verified that this 
choice also leads to vanishing scalar modes. Another possibility would 
be not to perturb the brane positions at all. This then leaves two fewer 
degrees of freedom, together with two fewer equations (i.e. Eq.~(\ref{juns3}) 
evaluated at the two branes). Again we have verified that in this case 
the remaining equations are not degenerate, so that no scalar modes 
are present.

In summary, all the inequivalent choices for the brane mass terms that 
we have considered lead to vanishing scalar modes. Although these choices
appear as the most natural possibilities, clearly we have not exhausted 
all the possible choices. Hence, we cannot rule out the possibility that 
some other brane term can lead to nonvanishing modes also in the scalar sector.

\section{Mode properties}

\subsection{Localization of the zero modes}

The existence of zero modes follows from the relation (\ref{alpha})
between the bulk and boundary mass parameters. However, this
relation only fixes one of the parameters, say $a$, so that there is 
still freedom to choose $\alpha$. Since the wavefunction depends
on $\alpha$ we can arbitrarily localize the zero modes anywhere in the bulk.
This is similar to previous studies~\cite{gp1} involving fields with 
spin $< 2$. Consider the effective four-dimensional action for the
tensor modes. In terms of the RS variables $y$ we obtain:
\begin{equation}
    S_{eff}= -\frac{1}{4} M^3 N_T^2 \int d^4x \int_0^{\pi R} dy\,
      e^{-2(1-4\alpha)k y} \,\partial^\rho H_{\mu\nu}^{(0)}(x)
       \partial_\rho H^{(0)\,\mu\nu}(x) + \dots~,
\label{teffaction}
\end{equation}
where we have used the solution (\ref{t0mode}) in the effective bulk action at 
quadratic order (see Appendix~\ref{appB}; note that indices in 
(\ref{teffaction}) and (\ref{veffaction}) are raised with $\eta_{\mu\nu}$). 
From (\ref{teffaction}) we see that the tensor zero mode wavefunction 
$f_T^{(0)}$ has a $y$ dependence:
\begin{equation}
       f_T^{(0)}(y) \propto e^{-(1-4\alpha)k y}~.
\label{t0wavefn}
\end{equation}
When $\alpha=0$ we obtain the RS tensor mode localized
on the UV brane. However we now see that by varying $\alpha$
we can smoothly deform the tensor mode to be localized anywhere. 
For $\alpha<0$ the mode becomes even more localized on the UV brane,
compared to the original RS scenario. On the
other hand, for $\alpha>0$ the tensor mode is delocalized away from the
UV brane towards the IR brane. The transition occurs for $\alpha=1/4$
where the tensor mode is completely flat. As $\alpha$ becomes larger than 
$1/4$ the tensor mode becomes more and more localized on the IR brane.
Hence, we have a continuous deformation of the original RS 
tensor mode from being completely localized on the UV brane
to being completely localized on the IR brane!

Similarly for the vector mode we obtain the effective action:
\begin{equation}
    S_{eff}= -\frac{1}{4} M^3 N_V^2 \int d^4x \int_0^{\pi R} dy\,
      e^{-4(1-2\alpha)k y} \,F_{\mu\nu}^{(0)}(x)F^{(0)\,\mu\nu}(x) + \dots~,
\label{veffaction}
\end{equation}
where we have used (\ref{newvec}) in the quadratic bulk action 
and $F_{\mu\nu}^{(0)}(x) =\partial_\mu 
b_\nu^{(0)} (x)-\partial_\nu b_\mu^{(0)}(x)$. Therefore, the vector zero 
mode has the wavefunction dependence: 
\begin{equation}
       f_V^{(0)}(y) \propto e^{-2(1-2\alpha)k y}~.
\label{v0wavefn}
\end{equation}
Again we see that the vector mode can be localized anywhere in the bulk.
In particular when $\alpha=1/2$ the vector mode is flat. Note that
there is a difference between the tensor and vector mode wavefunction 
dependence so that these modes are never localized at the same place.

\subsection{4D Planck mass}

The 4D (reduced) Planck mass $(M_P)$ is defined by assuming that there 
is matter on the branes described by an energy-momentum tensor 
$T_{\mu\nu}^{(i)}$:
\begin{eqnarray}
\label{mplanck}
S_{matter}&=&\int d^4x \, N_T\,A^{-4\alpha}(z_i)\,T^{(i)\,\mu\nu} 
       H_{\mu\nu}^{(0)}(x)~, \nonumber \\ 
&\equiv&\int d^4x \, \frac{1}{M_P}T^{(i)\,\mu\nu} 
       {\widetilde H}_{\mu\nu}^{(0)}(x)~,
\end{eqnarray}
where ${\widetilde H}_{\mu\nu}^{(0)}$(x) is the canonically normalized 
tensor zero mode. Thus, the four dimensional Planck mass is:
\begin{equation}
     M_P^2= \frac{M^3}{k} \frac{A^{8\alpha}(z_i)}{(4\alpha-1)}
      \left[ e^{2(4\alpha-1)\pi k R}-1\right]~.
\label{mpl}
\end{equation}
When $\alpha=0$ this expression reduces to the RS result~\cite{rs}:
\begin{equation}
        M_P^2\simeq \frac{M^3}{k}~,
\end{equation}
assuming $\pi k R \gg 1$. In particular notice that the RS
result does not depend on where the matter is localized.

However, when $\alpha\neq 0$ the Planck mass depends on whether matter 
is localized on the UV brane or IR brane. For matter on the UV brane, 
$A(z_0)=1$, and we obtain:
\begin{equation}
     M_P^2= \frac{M^3}{k(4\alpha-1)}
      \left[ e^{2(4\alpha-1)\pi k R}-1\right]\simeq
\begin{cases}
   \frac{M^3}{k(4\alpha-1)}\,e^{2 (4\alpha-1)\pi k R}\,
& \quad \alpha>\frac{1}{4}~,\\
M^3\, 2\pi R\,
& \quad \alpha=\frac{1}{4}~,\\
\frac{M^3}{k(1-4\alpha)}
& \quad \alpha<\frac{1}{4}~,\\
\end{cases}
\label{mpuv}
\end{equation}
where we have assumed $\pi k R\gg 1$ in the approximate expressions.
These expressions are consistent with the behavior of the graviton 
wavefunction. In particular when $\alpha <1/4$ the Planck mass
is consistent with the fact that the graviton wavefunction is localized on
the UV brane and there is no exponential expression.
The $\alpha=1/4$ result follows from the fact that in this limit the 
graviton wavefunction is flat and we recover the flat space 
result~\cite{add}.
For $\alpha > 1/4$ the graviton is localized towards the IR brane
and the coupling to matter in the UV brane is exponentially suppressed.

Instead when matter is localized on the IR brane, $A(z_1)=e^{-\pi k R}$,
we obtain:
\begin{equation}
          M_P^2= \frac{M^3}{k(4\alpha-1)}
      \left[ e^{-2\pi k R}-e^{-8\alpha\,\pi k R}\right]\simeq
\begin{cases}
   \frac{M^3}{k(4\alpha-1)}\, e^{-2\pi k R}\,
& \quad \alpha>\frac{1}{4}~,\\
M^3\, 2\pi R\, e^{-2\pi k R}\,
& \quad \alpha=\frac{1}{4}~,\\
\frac{M^3}{k(1-4\alpha)}\, e^{-8\alpha\pi k R}
& \quad \alpha<\frac{1}{4}~,
\end{cases}
\end{equation}
again assuming $\pi k R\gg 1$ in the approximate expressions. These 
expressions are consistent with the localization properties of the 
graviton, except that now there is a nontrivial warp factor for
the matter on the IR brane.

An interesting phenomenological scenario occurs when the graviton is
localized on the IR brane and all the standard model matter is located
on the UV brane. If we associate the IR brane with the millimeter scale
$10^{-3}$ eV, then for $M\sim k\sim$ TeV, we obtain the 
usual Planck mass $M_P$ for $\alpha=1/2$. The scale of the UV brane is the
TeV scale. In the bulk the tensor mode has a profile $e^{k y}$, and it is 
therefore localized away from the UV brane.
This explains the weakness of gravity with respect to the gauge interactions.
One can also verify that the same exponential suppression also holds for 
the coupling of the massive tensor modes. The vector zero mode is instead 
flat in the bulk. However, the vector mode does not couple to brane sources 
with a conserved energy-momentum tensor. Hence, conserved matter on the brane
interacts only with the graviton tensor mode.

We will see that the dual picture is particularly interesting since it 
indicates that
the graviton is composite at the millimeter scale, and emerges from the
dual CFT as a massless bound state. This emergent gravity picture is
not in contradiction with the Weinberg-Witten theorem~\cite{ww} because 
as in models of induced gravity, there is a source gravitational 
background which invalidates the assumptions of the theorem. 
The corresponding energy-momentum tensor is no longer conserved and
can obtain a (large) anomalous dimension. Since the UV scale is associated 
with the TeV scale the gauge hierarchy problem is trivially solved (as usual, 
additional dynamics may be necessary to generate and stabilize the IR scale).

\subsection{Short range modifications of gravity}

We are interested in the gravitational interaction between
nonrelativistic matter sources on the UV brane. This interaction is due 
to the tensor perturbation of the geometry. It includes the contributions 
from both the zero mode (already studied in  Section 4.2 above) and the 
massive Kaluza-Klein modes. As we have seen in~(\ref{tenKKmass}), the 
lowest Kaluza-Klein mass is at the IR scale 
$m_1 \sim k \, A \left( z_1 \right) \,$. This scale sets the
distance above which gravity is standard. We are interested in the 
gravitational interactions at distances $r$ smaller than this, but still 
much greater than $1/k$ (so that they can be phenomenologically relevant):
\begin{equation}
\frac{1}{k} \ll r \ll \frac{1}{k \, A_1} \,\,,
\label{intermediate}
\end{equation}
where we have denoted $A_1 \equiv A \left( z_1 \right) \,$. The computation 
is given in Appendix D. In the following we present the final result and 
discuss it.

The contribution from the Kaluza-Klein modes presents two distinct behaviors 
depending on whether $\alpha$ is smaller or greater than $1/4 \,$. For either 
region, we define a positive parameter $\xi\equiv\vert 4 \alpha - 1 \vert \,$ 
(therefore, the two regions are joined at $\xi = 0 \,$; notice however that 
we have chosen $\xi$ to be positive in both of them). The gravitational 
potential is found to be:
\begin{eqnarray}
V(r) &\simeq& - \frac{\mu}{M_P^2 \, r} \left[ 1 + \frac{2 \, \Gamma 
\left( 2 \xi \right)}{\xi \, \Gamma^2 \left( \xi \right)} \, \frac{1}
{\left( 2 \, k \, r \right)^{2 \xi}} \right] \;\;\;,\;\;\; \xi = 1 - 4 \, 
\alpha \,\,,\,\, \alpha < 1/4 \,\,, 
\label{grapot1} \\
V(r) &\simeq& 
- \frac{\mu}{M_P^2 \, r}\times\begin{cases}
1 + \frac{2 \, \Gamma \left( 2 \xi \right)}
{\xi \, \Gamma^2 \left( \xi \right)} \, \frac{1}{\left( 2 \, k \, A_1 \, r 
\right)^{2 \xi}}~~~,& r \gappeq \frac{1}{k A_1}\\[5mm]
\frac{2 \, \Gamma \left( 2 \xi \right)}
{\xi \, \Gamma^2 \left( \xi \right)} \, \frac{1}{\left( 2 \, k \, A_1 \, r 
\right)^{2 \xi}}~~~,& r \lappeq \frac{1}{k A_1}\\
\end{cases}
;\xi = 4\alpha - 1,\alpha > 1/4.
\label{grapot2}
\end{eqnarray}
The $1/r$ term in these two expressions is the contribution from the zero 
mode, which reproduces the standard Newtonian gravity at large distances. 
Instead, the second term represents the interaction mediated by the 
Kaluza-Klein massive modes in each case. 

Note that  there is a striking phenomenological difference between the two 
cases, shown in (\ref{grapot1}) and (\ref{grapot2}), respectively. 
Indeed, in the first case $\alpha < 1/4$, the 
corrections to the Newtonian potential are relevant only at distances 
$r \lappeq \, k^{-1} \,$, that is only at the UV scale.\footnote{The RS 
model, characterized by $\alpha = 0 \,$ belongs to this interval; 
$\alpha = 0$ corresponds to $\xi = 1 \,$, so that we recover the $1/r^3$ 
correction computed in~\cite{rs2}.} However, in the second case of $\alpha > 
1/4 \,$, the gravitational potential is strongly modified already at the much 
larger (and, possibly, phenomenologically relevant) IR scale, $r \simeq 
\left( A_1 \, k \right)^{-1} \,$. Moreover, below this distance the $1/r$ 
term disappears and the potential is given solely by a power law 
$1/r^{2\xi+1}$. The cancellation of the $1/r$ term will be explained in the 
next section, but it is already clear that the corresponding force is stronger 
than the usual gravity. So, for $\alpha >1/4$ standard gravity only emerges 
at infrared distance scales $r \gappeq \left( A_1 \, k \right)^{-1}$.

Actually, this behavior is due to the different localization of the zero 
mode in the two regimes.\footnote{We do not expect the potential to be 
discontinuous at $\alpha = 1/4  \,$.  However, the two 
expressions~(\ref{grapot1}) and~(\ref{grapot2}) become invalid for 
$\xi \rightarrow 0$, since the approximations adopted to derive them break 
down in this limit. We expect that in this region the exact potential 
quickly (but smoothly) interpolates between the values given on the two 
sides.} For $\alpha < 1/4 \,$, the zero mode is localized towards the 
UV brane, and the relative contribution from the Kaluza-Klein modes can be 
neglected. On the contrary, for $\alpha > 1/4$ the zero mode is localized 
towards the IR brane  and the relative contribution of the massive modes 
significantly increases becoming the dominant contribution for 
$r \lappeq \left( A_1 \, k \right)^{-1}$---the regime indicated in 
(\ref{intermediate}). We will see that for $\alpha>1/4$ 
this behavior is consistent with the graviton being a composite
at the IR scale.

Let us briefly examine the experimental consequences of our composite
graviton model in the intermediate regime (\ref{intermediate}).  
Assuming that the IR (or compositeness) scale
is related to the cosmological constant then the IR scale is 
$\sim 10^{-3}$ eV. As we have seen above at energies smaller 
than this scale we have the usual Newton law. The striking experimental 
signal of our model would be that Newton's law of gravity changes to a 
new power law $r^{-1} \rightarrow r^{- 2 \xi -1}$  below $\sim 0.1$ mm,
which could be fractional or even irrational!

\subsection{The composite graviton and Green's function analysis}

The modifications of Newton's law and the compositeness of the graviton
can also be studied by computing the 5D bulk Green's function associated 
with the tensor mode $\widehat h_{\mu\nu}$. The formalism for computing 
5D propagators
in a slice of AdS can be found in Appendix A of Ref.~\cite{gp2}. The 
Lorentz index structure of $\widehat h_{\mu\nu}$ can be neglected, and then
the Green's function of the tensor mode is identical to that of a bulk 
scalar mode with mass $\widehat M_\Phi^2 = 4a k^2$ and boundary condition
$r=4\alpha_\pm$ (using the notation of Ref.~\cite{gp2}). A straightforward
substitution into the general expressions in Ref.~\cite{gp2} then gives rise
to the Planck brane-Planck brane Green's function:
\begin{equation}
  G(x,z_0; x',z_0)=\int \frac{d^4p}{(2\pi)^4}\, e^{ip\cdot(x-x')}\, 
      G_p(z_0,z_0)~,
\end{equation}
where: 
\begin{equation}
     G_p(z_0,z_0)=\pm\frac{2 \nu k}{p^2} -\frac{1}{p}
       \left[\frac{I_\nu(p L)K_{\nu\pm 1}(p/k)+ K_\nu(p L) I_{\nu\pm 1}(p/k)} 
{I_\nu(p L)K_\nu(p/k) - K_\nu(p L) I_\nu(p/k)}\right]~.
\label{greenfn}
\end{equation}
Here $p \equiv \sqrt{p^2}$,  $\nu=\pm(4\alpha_\pm -1)\equiv\nu_\pm$, $L^{-1} = kA(z_1)=k 
e^{-\pi kR}$ is the IR scale, and $I_\nu(z), K_\nu(z)$ are the modified 
Bessel functions. 
The Green's function (\ref{greenfn}) has been written as a sum of two 
terms~\cite{gkr}. The first term represents a part of the zero mode 
contribution, while the more complicated second term receives contributions 
from both the zero mode and the Kaluza-Klein
tower of massive modes: the zeros of the denominator of the second term at 
$p = i m_n$ precisely reproduce  the Kaluza-Klein mass spectrum 
(\ref{tenKKmass}). Moreover, the contribution from the zero mode is clear
from the fact that also this term 
scales as $1/p^2$ for $p \rightarrow 0$.

The contribution of the massive states can be simplified by expanding 
the Green's function in the 4D momentum $p$. Since $k$ is the AdS curvature
scale we will assume $p\ll k$. Consider first the case of $\alpha<1/4$. 
This corresponds to the localization of the graviton on the UV brane.
In the limit of either $pL \gg 1 $ or $pL \ll 1$ the dominant 
contribution to the Green's function is always:
\begin{equation}
\label{exp1} 
     G_p(z_0,z_0) \simeq 2(4\alpha-1)\frac{k}{p^2} + \dots 
       \simeq -\frac{2 M^3}{M_P^2}\frac{1}{p^2} + \dots~,
\end{equation}
where in the second equality we have substituted for the Planck mass from 
(\ref{mpuv}). Thus, we see that at low energies $pL \ll 1$ only the zero mode
contributes to the Green's function and the massive modes are decoupled.
Similarly, for $pL\gg 1$ the zero mode is again dominant. This behavior
is consistent with the gravitational potential (\ref{grapot1}), and the
fact that the tensor zero mode is localized on the UV brane and remains 
pointlike for all energies $p\ll k$.

Let us now consider the case where the graviton is localized on the
IR brane. Assuming $\alpha > 1/4$, we find that in the limit $pL \ll 1$:
\begin{equation}
\label{exp2}
     G_p(z_0,z_0) \simeq -2\xi e^{-2\xi\pi k R}\;       
       \frac{k}{p^2} + \dots \simeq 
    -\frac{2 M^3}{M_P^2}\; \frac{1}{p^2} + \dots~,
\end{equation}
where $\xi = 4\alpha -1$ and the Planck mass has been substituted from 
(\ref{mpuv}). Again we see that the zero mode gives the dominant contribution 
at energies below the IR scale. There is now also an extra volume 
suppression that is absorbed into the gravitational coupling. 
On the other hand at energies above the IR scale, when
$pL \gg 1$, we obtain:\footnote{This expansion assumes $\xi \neq1/2 \,$.
For $\xi = 1/2 \,$, the analytic terms are not present in the series
expansion of~(\ref{exp3}).}
\begin{eqnarray}
\label{exp3}
     G_p(z_0,z_0) &\simeq& -\frac{1}{p} \frac{K_{\xi-1}(p/k)}
    {K_{\xi}(p/k)}\\
    &\simeq& \frac{1}{2(1-\xi)k} \left[1-\frac{p^2}{4 k^2 
    (\xi-1)(\xi-2)} -\left(\frac{p}{2k}\right)^{2\xi-2} 
    \frac{\Gamma(2-\xi)}{\Gamma(\xi)}+\dots \right]~,\nonumber
\end{eqnarray}
Remarkably, there is now no longer a dominant zero mode contribution since
the massive Kaluza-Klein modes cancel the leading $1/p^2$ contribution in 
(\ref{greenfn})! We will see that this behavior is consistent with the 
fact that the graviton is composite at the IR scale $L^{-1}$. When $pL \ll 1$ 
the graviton is essentially pointlike and contributes in the usual way, 
but when $p L \gg 1$ the zero mode effectively disappears. Thus 
standard gravity only appears at energies below the compositeness scale 
$L^{-1}$ and is seen to emerge from the CFT, whereas above the compositeness 
scale (or $r \lappeq L$) the dominant CFT contribution changes Newton's law 
to a different power law~(\ref{grapot2}), which can be fractional and
even irrational.

The cancellation of the leading $1/p^2$ term can also be seen in coordinate 
space for special values of $\alpha$ where the sum over Kaluza-Klein modes
can be done analytically. Consider the case of $\nu_-=-1/2$ or $\alpha=3/8$ 
(recall that $\alpha>1/4$ corresponds to IR-brane localization).
The Green's function is:
\begin{eqnarray}
    G_p(z_0,z_0) = -\frac{1}{p}\coth(p z_1)~
                 = -\frac{1}{z_1}\left[\frac{1}{p^2}
         +2\sum_{n=1}^{\infty} \frac{1}{p^2+m_n^2}\right]~,
\label{gfexactm}
\end{eqnarray}
where $m_n=\pi n/z_1$ (note that this, in fact,
 is the Green's function for an even field on a flat $S^1/Z_2$ orbifold of size $z_1$, evaluated at one of the fixed points). 
It is straightforward to check  that
the $1/p^2$ pole is cancelled for $p z_1\gg 1$. Alternatively the 
gravitational potential can be calculated exactly using the expressions 
(\ref{summodes}) and (\ref{exactexp}) in  Appendix D. One obtains:
\begin{eqnarray}
    V(r) = -\frac{\mu}{M_P^2}\frac{1}{r}\left[1+2\sum_{n=1}^{\infty} 
    e^{-m_n r}\right]~
      = -\frac{\mu}{M_P^2}\frac{1}{r} \coth\left(\frac{\pi r}{2z_1}\right)~.
\end{eqnarray}
Note that the Fourier transform of $V$ is proportional to the
Green's function (\ref{gfexactm}).
For distances $r<z_1 \simeq L$ the series expansion gives:
\begin{equation}
     V(r)\simeq -\frac{\mu}{M_P^2}\frac{2 z_1}{\pi r^2} +\dots~.
\end{equation}
This agrees with the gravitational potential (\ref{grapot2}) for 
$r< L=1/(kA_1)$. 

For $\alpha<1/4$, one can also analytically verify that 
the $1/r$ term does not cancel when $\nu_-=1/2$ (or $\alpha =1/8$). 
This is obvious because in this case the exact Green's function is:
\begin{equation}
    G_p(z_0,z_0) = -\frac{1}{p^2} - \frac{1}{p}\coth(p z_1)~,\\
\end{equation}
and the extra $-1/p^2$ term compared to (\ref{gfexactm}), corresponding
to the $1/r$ term in the potential, is no longer cancelled at high energies.

These two simple examples enlighten our gravitational potential calculation
for $\alpha>1/4$, and reveal that when the graviton is localized on the IR
brane the usual Newton's law of gravity only emerges as 
a low-energy phenomenon. Remarkably, as we will see in the next section, 
these properties can also be described purely in a 4D dual interpretation.

\section{The holographic interpretation}

Motivated by the string theory AdS/CFT correspondence, bulk theories in a 
slice of AdS$_5$ can be given a holographic interpretation as dual to a CFT 
(at large-$N$ and large 't Hooft coupling) with conformal invariance broken 
in the IR, coupled to gravity and possibly other fields~\cite{pheno1,
pheno2,pheno3}. As opposed to the string-theory AdS/CFT, the UV boundary 
value of bulk fields are not only sources of operators in the dual CFT but 
acquire their own dynamics due to the presence of a UV cutoff.

The nature of the holographic interpretation of bulk theories in a 
slice of AdS$_5$ is necessarily speculative; however, it passes some quantitative tests, despite the fact that one usually does not  
know what the dual CFT is, or even whether it actually exists! Nevertheless, 
in most cases  one is able to give a reasonably convincing  picture of the 
dual 4D strong dynamics, which may be useful as a guide to finding a theory 
with the desired features. With these caveats in mind, in this section we 
attempt a quantitative dual 4D description of the mass-deformed gravity in 
a slice of AdS$_5$ studied in the previous sections.

The study of the 4D dual begins by recalling that the UV-boundary values of the bulk fields are sources of operators in the dual CFT. The classical bulk action, evaluated on solutions of the bulk equations of motion with arbitrary values on the UV brane (and obeying the proper boundary conditions on the IR brane) is interpreted by the stringy AdS/CFT correspondence as being equal to the generating functional of connected Green's functions of the 4D dual CFT.  To continue, we begin with the solution of the bulk equation which obeys the boundary condition on the IR brane:  
\begin{equation}
\label{bulkIRsoltn}
{\widehat H}(p,z) = {\widehat H}(p) A^{-2}(z) \left( J_{\nu_\pm \mp 1}(iq) - 
Y_{\nu_\pm \mp1} (iq)\; \frac{J_{\nu_\pm}(i q_1)}{Y_{\nu_\pm}(i q_1)}
\right)~,
\end{equation}
where $\nu_\pm = \pm (4 \alpha_\pm - 1)$, $q = \frac{p}{k A(z)}$, $q_1 = 
\frac{p}{k A_1}$; $q_0 = \frac{p}{k A_0}$; $p^2 =- m^2$ is the mass-shell 
condition,  and ${\widehat H}(p,z)$ is the 4D Fourier transform of $\widehat{h}(x,z)$.
Note that we will omit tensor indices, as we are only concerned here with 
the transverse-traceless components of the metric perturbation, which obey 
a scalar equation. The bulk action, which generates Green's functions in the 
dual is:
\begin{equation}
S_{bulk} = \frac{M^3}{4} \int \frac{d^4 p}{(2 \pi)^4} \left[ A^3\;  
{\widehat H}(p,z) ({\widehat H}^\prime(-p,z) - 4 \alpha A k 
{\widehat H}(-p,z))\right]\bigg\vert_{z = z_0}~,
\end{equation}
evaluated on the bulk solution (\ref{bulkIRsoltn}). This can be rewritten as:
\begin{equation}
S_{bulk} = \frac{M^3 k}{4} \int \frac{d^4 p}{(2 \pi)^4}  \; F(q_0, q_1) \; 
{\widehat H}(p) {\widehat H}(-p)~, 
\end{equation}
where:
\begin{equation}
F(q_0, q_1) = \mp iq_0 \left[ J_{\nu_\pm \mp 1}(iq_0) - Y_{\nu_\pm \mp1} (iq_0)\; \frac{J_{\nu_\pm}(iq_1)}{Y_{\nu_\pm}(iq_1)} \right] 
\left[  J_{\nu_\pm}(iq_0) - Y_{\nu_\pm} (iq_0)\; 
\frac{J_{\nu_\pm}(iq_1)}{Y_{\nu_\pm}(iq_1)} \right]~.
\end{equation}
The dual theory two point function of the operator ${\cal O}$ sourced by the bulk field ${\widehat H}$, $   \langle {\cal O}  {\cal O} \rangle(p) \equiv 
\int d^4 x~e^{- i p\cdot x} \langle T {\cal O}(x) {\cal O}(0) \rangle$, is 
contained---up to local counterterms which we will discuss later---in the 
second derivative, $\Sigma(p)$, of $S_{bulk}$ with respect to the boundary 
value of the metric perturbation $A_0^2 \widehat{h}$. 
The correlator is, in various equivalent forms to be used later: 
\begin{eqnarray}
  \Sigma (p) &=& \int d^4 x e^{- i p\cdot x} \frac{\delta^2 S_{bulk}}
{\delta (A_0^2 \widehat{h}(x, z_0)) \delta (A_0^2 \widehat{h} (0, z_0))}~
\nonumber \\
&=&\left(\frac{M}{k}\right)^3 \; \frac{k^4}{2} \; (\mp iq_0) \: 
\frac{J_{\nu_\pm} (iq_0) Y_{\nu_\pm}(iq_1) - Y_{\nu_\pm}(iq_0) 
J_{\nu_\pm} (iq_1)}{J_{\nu_\pm \mp 1} (iq_0) Y_{\nu_\pm}(iq_1) 
- Y_{\nu_\pm \mp 1}(iq_0) J_{\nu_\pm} (iq_1) }\nonumber~\\
&=& \left(\frac{M}{k}\right)^3 \;\frac{k^4}{2} \frac{q_0 \left( I_\nu(q_0) 
K_\nu(q_1) - I_\nu(q_1) K_\nu(q_0) \right)}{I_{\nu \mp 1}(q_0) K_\nu(q_1) 
+ I_\nu(q_1) K_{\nu \mp 1}(q_0)}\nonumber \\
&=& \frac{M^3}{2 A_0^4} \frac{1}{G_{p}(z_0, z_0)}~,
\label{correlator}
\end{eqnarray}
where $G_{p}$ is the boundary-to-boundary propagator of (\ref{greenfn})  divided by $A_0^3$ and with $k$ replaced by $A_0 k$.

The behavior of $\Sigma(p)$ can be studied for momenta $p$ such that 
$kA_0 \gg p \gg  k A_1$, or equivalently, $q_0 \ll 1, q_1 \gg 1$. In 
this energy regime, the effects of the conformal symmetry breaking 
(i.e., the IR brane) are completely negligible. The leading nonanalytic 
piece in  $\Sigma(p)$ is then interpreted, by ``matching'' to the string 
AdS/CFT correspondence in the $A_0 \rightarrow \infty$ limit, as due to the 
strong dynamics of the dual CFT above the scale of conformal symmetry 
breaking. On the other hand,  the analytic pieces in the correlator, which, 
in  string  AdS/CFT, are subtracted away by adding appropriate counterterms, 
are now interpreted as kinetic (and higher-derivative terms) of the dynamical 
source field in the holographic dual. 

\subsection{$\alpha_-$ branch holography}

We will first consider the $\alpha_-$ branch of our solution, which is the one continuously connected to the $\alpha = 0$ RS value (see Figure~\ref{fig1}).
For  $\frac{1}{2} > \alpha_- >0$, or  $-1 < \nu_- < 1$, we 
find:\footnote{Notice that the expression (\ref{series1}) does not 
contradict (\ref{exp3}) since when $0<\xi<1$ (or $-1<\nu_-<0$) the 
nonanalytic term dominates in (\ref{exp3}), so that the constant term 
disappears in the inverse. Also, for $\nu_-=-1/2$ there is no analytic term
in the series expansion in this regime.}
\begin{eqnarray}
\label{series1}
  \Sigma(p) \simeq 
- \left( \frac{M}{k}\right)^3 \frac{k^4}{2} \left(  \frac{q_0^2}{2\nu}   
+ q_0^{2 \nu + 2}\; \frac{\Gamma(-\nu)}{2^{2 \nu + 1} \nu\Gamma(\nu)} 
+...\right)~,
\end{eqnarray}
while for all other values on the $\alpha_-$-branch ($\alpha_- < 0$ or $\nu = \nu_- > 1$):
\begin{eqnarray}
\label{series2}
\Sigma(p) \simeq   - \left( \frac{M}{k}\right) ^3 \frac{k^4}{2} \left[ 
\frac{q_0^2}{2\nu} \left(1 + ... + c q_0^{[2 \nu]}\right) + q_0^{2 \nu + 2}
\;\frac{\Gamma(-\nu)}{2^{2 \nu + 1} \nu \Gamma(\nu)}   + ... \right],
\end{eqnarray}
where $[2\nu]$ denotes the largest integer smaller than $2 \nu$. 
In each case (\ref{series1}), (\ref{series2}), we have included the leading analytic piece as well as all terms up to the leading nonanalytic piece, $q_0^{2 \nu + 2}$. The power of $q_0$ in the nonanalytic piece indicates that the scaling dimension $\Delta_{\cal O}$ of the operator ${\cal O}$---the energy momentum tensor of the dual theory---sourced by the metric perturbation $h$ is:
\begin{equation}
\label{dimO}
\Delta_{\cal O} = 3 + \nu = 4 - 4 \alpha_-~,
\end{equation}
on the $\alpha_-$ branch. The leading analytic piece in (\ref{series1}), 
(\ref{series2}) indicates that there is a kinetic term for the metric 
perturbation in the dual theory. 

\begin{figure}
\centerline{
\includegraphics[width=1.0\textwidth]{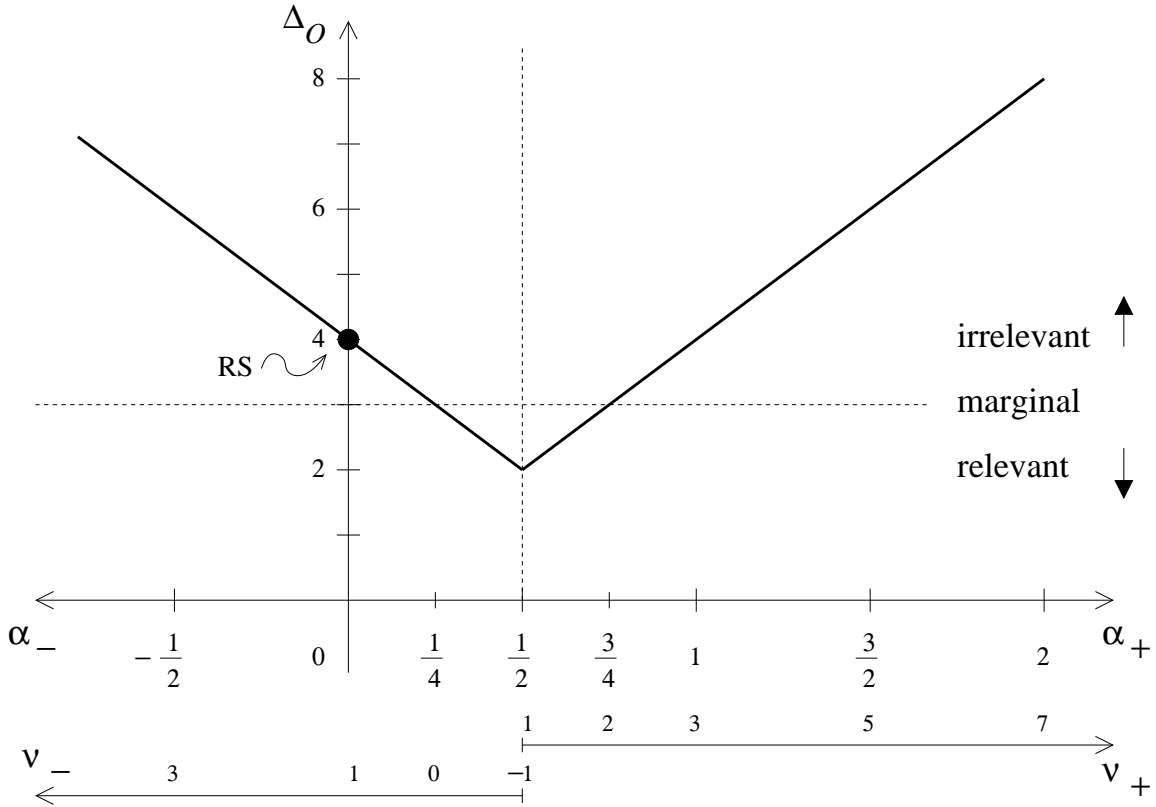}
}
\caption{The scaling dimension $\Delta_{\cal O}$ plotted as a function of
the boundary mass parameter $\alpha$. The source coupling to the CFT is 
irrelevant for $\alpha< 1/4$ and $\alpha>3/4$, marginal for $\alpha=1/4$ and
$\alpha=3/4$ and relevant for $1/4 < \alpha < 3/4$. Note also that the 
RS value at $\alpha=0$, and the specular RS value at $\alpha=1$ both have
$\Delta_{\cal O}=4$.
}
\label{dimOfig}
\end{figure}

Thus, the holographic description of this branch is that of a metric fluctuation  $\widehat{h}_{\mu\nu}$ coupled to  $T^{\mu\nu}_{CFT}$ of scaling dimension $4 - 4 \alpha_-$. Note  the unusual fact that the energy momentum tensor of the CFT has an anomalous dimension. This, however, is required if the 4D dual is to evade the Weinberg-Witten theorem---which assumes Poincare invariance, broken here by the presence of a nontrivial background metric, as in theories of induced gravity.  Notice also, from (\ref{dimO}), that the scaling dimension of 
$T^{\mu\nu}_{CFT}$ can be as low as 2, for  $1/2 > \alpha _- > 0$, as can
be seen in Fig.~\ref{dimOfig}. A scaling behavior of the CFT with such 
a large anomalous dimension should persist no matter how small the breaking 
of Poincare invariance (or, equivalently, the deviation of the metric 
background from Minkowskian)! Leaving aside the issue of existence CFTs 
with such behavior, we continue with our attempt at giving a  
(semi-)quantitative picture of the dual dynamics.

\subsubsection{The dual theory and its dynamics}

The Lagrangian of our dual theory is, then,  at a UV scale $ \sim k$, with a canonically  normalized metric perturbation: 
\begin{equation}
\label{UVlagrangian}
{\cal L}_{UV}  =   \epsilon_\nu \frac{1}{4}\,h_{\mu\rho} \Box h^{\mu\rho} 
+ \frac{\lambda_{UV}}{k}\;  h_{\mu\rho}  T^{\mu\rho}_{CFT}  +
\frac{\lambda_{UV}}{k}   \; h_{\mu\rho}  T^{\mu\rho}_{matter}~ 
+ {\cal L}_{CFT}~, 
\end{equation}
where $ \epsilon_\nu = {\rm sign} \; \nu$ and 
$\lambda_{UV} = |\nu|^{1/2} (M/k)^{-3/2}$. We have included the 
coupling to observable matter fields (UV-brane localized in the gravity 
dual).\footnote{Note that in the  case of irrelevant coupling, the sign of the kinetic term for $h_{\mu\nu}$ is the proper one, as  $\epsilon_\nu = 1$ for $\alpha_- < 1/4$; while it has the wrong sign in the case of relevant coupling; we will see below that the leading contribution to the $h_{\mu\rho}$ kinetic term of the right sign arises from IR physics not accounted for in (\ref{UVlagrangian}). }

From eqn.~(\ref{UVlagrangian}), taking into account the anomalous scaling dimension of $T_{CFT}$ from (\ref{dimO}), we conclude that 
the coupling of the metric perturbation to the  CFT  energy momentum tensor is relevant for $\alpha_- > 1/4$, marginal if $\alpha_- = 1/4$, and irrelevant for $\alpha_- < 1/4$. Introducing a renormalization scale $\mu$, the dimensionless coupling is then $\lambda(\mu) \equiv (\mu/k)^{1- 4 \alpha_-}  \lambda_{UV} $ and satisfies the RGE:
\begin{equation}
\label{RGE1}
\mu \frac{d \lambda}{d\mu} = - (4 \alpha_- - 1) \lambda + \ldots
\end{equation}
where the first term is a result of dimensional analysis and 
higher order terms due to the  CFT's  interactions have been neglected. 

If  $\lambda$ is relevant, i.e. $\alpha_- > 1/4$, the solution of the RGE:
\begin{equation}
\label{relevantsolution}
\lambda_{IR} = \lambda_{UV} \left( \frac{k}{m_{IR}} \right)^{4 \alpha_- - 1}~, 
\end{equation}
indicates that the coupling of $h_{\mu\nu}$ to the CFT is enhanced in the IR. 
The conformal invariance is broken at the IR scale $m_{IR}$ and we expect 
that integrating out the CFT dynamics at the IR scale will induce a kinetic 
term for $h_{\mu\nu}$. From the CFT point of view 
the dynamics is both strong and unknown; however,  it is 
clear (see, e.g., \cite{Adler:1982ri}) that producing a kinetic term requires two insertions of the 
dimensionless coupling $\lambda_{IR}$, as indicated in eqn.~(\ref{IRlagrangian}) below. 

Clearly, in the weakly coupled
gravity dual we can directly compute this contribution by calculating the two point function 
$\Sigma$ in the IR limit $p\ll k A_1$. The pure IR 
contribution is obtained by subtracting the analytic piece of 
$\Sigma$ that arose 
in the limit $p\gg k A_1$, see (\ref{series1}).  Thus, expanding (\ref{correlator}) for small $q_0$ and $q_1$  and subtracting 
the analytic term already accounted for in (\ref{series1}) (and in the kinetic term in (\ref{UVlagrangian})) leads to the  pure IR contribution to the correlator:
\begin{equation}
\label{IRseries1}
  \Sigma(p)_{IR} \simeq 
\left( \frac{M}{k}\right)^3 \frac{k^4}{2} \left( A_1^{2\nu}
\; \frac{q_0^2}{2\nu} +...\right)~,
\end{equation}
where $A_1=m_{IR}/k$. Thus, the effective Lagrangian describing the 
long-wavelength fluctuations of $h_{\mu\nu}$ and its coupling to the 
observable matter sector is given by:
\begin{equation}
\label{IRlagrangian}
{\cal L}_{IR} = \left[ \epsilon_\nu - \left(\frac{M}{k}\right)^3
\frac{\lambda_{IR}^2}{\nu}\right]
\frac{1}{4} h_{\mu\rho}\Box h^{\mu\rho} + \frac{\lambda_{UV}}{k} h_{\mu\rho} 
T^{\mu\rho}_{matter}~.
\end{equation} 
In the relevant case, where $-1< \nu_- <0$, our gravity dual calculation shows 
that the IR contribution to the kinetic term for $h_{\mu\rho}$ has the correct
sign and dominates over the ghost-like $\epsilon_\nu$ contribution. 
Hence, using (\ref{relevantsolution}) and canonically normalising (\ref{IRlagrangian}), the coupling of the observable matter to gravity at scales below 
$m_{IR}$ is:
\begin{equation}
\label{relevantMpl}
M_{P} = \sqrt{\frac{M^3}{k |\nu_-|}} \;  \frac{\lambda_{IR}}{\lambda_{UV}} 
= \sqrt{\frac{M^3}{k|\nu_-|}}  \;
\left(\frac{k}{m_{IR}} \right)^{4 \alpha_- -1}~.
\end{equation}
Upon identifying the ratio  of UV to IR scales with the warp factor, 
\begin{equation}
\label{scalesIdent}
  \frac{k}{m_{IR}} = e^{\pi k R}~,
\end{equation}
Eq.~(\ref{relevantMpl}) gives $M_{P}^2 =\frac{M^3}{|\nu_-| k}\; 
e^{2\pi k R (4 \alpha_- - 1)}$, in agreement with the gravity dual 
result (\ref{mpuv}).

Consider next the irrelevant case with $\nu_->0$. In this case the
coupling of the metric perturbation to the CFT is irrelevant so that
the induced contribution proportional to $\lambda_{IR}$ in 
(\ref{IRlagrangian}) is negligible. Thus, in (\ref{IRlagrangian}) the 
leading term proportional to $\epsilon_\nu$ dominates (and has the right 
sign!), leading to $M_P = k/\lambda_{UV}$, which precisely  equals   
the last line in Eq.~(\ref{mpuv}) as well as the RS result.

Finally,  consider the case where the coupling of CFT to the background 
metric is marginal $(\nu_-=0)$. We have to take into account higher
order terms in (\ref{RGE1}), which we can calculate using the weakly 
coupled gravity description. In the IR limit $p\ll k A_1$ we obtain:
\begin{equation}
\label{IRseries0}
 \Sigma(p)_{IR} =
-\left( \frac{M}{k}\right)^3 \frac{k^4}{2} \left( q_0^2 \log\frac{A_0}{A_1}+...
\right)~,
\end{equation}
so that the IR Lagrangian becomes: 
\begin{equation}
\label{margIRlag}
{\cal L}_{IR} = \left(\frac{M}{k}\right)^3 
\log\left(\frac{A_0}{A_1}\right)^2 \frac{1}{4} h_{\mu\rho}\Box h^{\mu\rho} 
+ \frac{1}{k} h_{\mu\rho} T^{\mu\rho}_{matter}~,
\end{equation}
where $A_0/A_1 = k/m_{IR}$.
Thus canonically normalising the kinetic term and using (\ref{scalesIdent}) leads to a Planck mass:
\begin{equation}
   M_P^2 = k^2 \left(\frac{M}{k}\right)^3 2\log\frac{A_0}{A_1} = M^3 \; 2\pi R~.
\end{equation}
This again agrees with the corresponding result in (\ref{mpuv}).

Thus, we have a rather unusual "theory," particularly in the case of a 
relevant coupling of the CFT to gravity. We have two important hierarchical scales of nature, 
for definiteness take them to be  meV  ($m_{IR}$) and  TeV  ($k$), as in our discussion after Eq.~(\ref{mpuv}). 
The infrared scale is presumably determined by some dynamical mechanism---unspecified both in the 5D gravity 
and the 4D CFT descriptions---from the UV scale. The TeV scale is the cutoff of the theory,  where a more fundamental description takes over.
Naturally, in the UV theory, a Newton constant $G_N$ of order TeV$^{-2}$ is 
expected (the first term in (\ref{UVlagrangian}), possibly including 
additional bare contributions). Along the RG flow to lower and lower scales, 
the hidden CFT sector becomes stronger and remains so until, at the meV  
scale,  conformal invariance is broken in the strongly coupled  CFT. Thus, 
the  meV  scale ``broken CFT'' induces a Newton constant.  It is strong enough
that despite the fact that it operates at  meV  scales, the induced Planck 
scale is  hierarchically larger: $M_P^2 \gg$ TeV$^2$! In other words, gravity 
is so weak because of the strength of the hidden CFT over a large interval of 
scales. 

While this ``scenario'' sounds really unusual, and we are not aware of a CFT with the desired properties,  it is not so difficult to come up with a weak coupling---in fact, free field theory---model of this phenomenon. Consider    $N$ free fields (of whatever nature, so long as they induce the correct sign Newton constant), coupled only to gravity, with a characteristic mass scale  $\sim$meV. Above the   meV  scale, this hidden theory is conformal, which protects the UV modes of these  fields from generating a $G_N$. But conformal invariance is broken at  the meV scale and so one expects a contribution to $M_P^2$ which will be of order $N\times$meV$^2$. Now if we take $N = M_P^2/$meV$^2$, then this clearly 
dominates the  TeV$^2$  contribution of ordinary massive matter.

As far as the evolution of the universe, our picture would predict that the 
strength of gravity should change  from TeV to $M_P$ during cosmological 
evolution; note that this does not affect BBN as the transition to ``normal'' 
strength gravity occurs before nucleosynthesis, when the Hubble size is of 
order meV$^{-1}$. There may, however, be relevant consequences for the physics 
of the earlier universe, as for instance inflation and 
baryogenesis.\footnote{It should be clear that we have nothing to say about 
the cosmological constant.}

\begin{figure}
\centerline{
\includegraphics[width=0.6\textwidth]{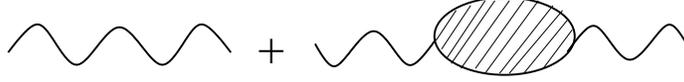}
}
\caption{The Feynman diagrams in the 4D dual theory responsible for the
gravitational potential corrections (which need to be summed up
for the case of a relevant coupling). The source field $h_{\mu\nu} \,$, 
indicated by a wavy line, interacts with the CFT contribution,
indicated by the blob.}
\label{cftfig}
\end{figure}

\subsubsection{The gravitational potential in the dual theory}

Let us now describe the leading correction to Newton's law at intermediate distances, $ r < 1/(k A_1)$, from the point of view of the dual interpretation. In the irrelevant case ($\nu_- > 0$), the coupling of matter to the CFT can be treated perturbatively and the leading correction arises from a single insertion of the CFT correlator and two insertions of the source field, as in the RS case (see Figure~\ref{cftfig}). The leading and first subleading contribution to the Newton potential is:
\begin{eqnarray}
\label{alphaminusirrelevant}
V(r) &=& -\mu \frac{\lambda_{UV}^2}{k^2}\int \frac{d^3 p}{2 \pi^2}~e^{i p x}\; \left(\frac{1}{p^2}-  \frac{\lambda_{UV}^2}{k^2} \frac{\langle {\cal O} 
{\cal O} \rangle (p)}{p^4}\right)~,\nonumber\\
&=& -\mu \frac{\lambda_{UV}^2}{k^2}\int \frac{d^3 p}{2 \pi^2}~ e^{ip\cdot x}  
\left( p^{-2} - p^{2 \nu -2} k^{-2 \nu}  \frac{\Gamma(-\nu)}{2^{2 \nu} 
\Gamma(\nu)} \right)~,
\end{eqnarray}
where:
\begin{equation}
   \langle {\cal O} {\cal O} \rangle (p) = - \left(\frac{M}{k}\right)^3 
    k^4 \left(\frac{p}{2k}\right)^{2 \nu + 2} \frac{4 \Gamma(-\nu)}
    { \nu \Gamma(\nu)}~,
\end{equation}
is the nonanalytic piece of $\Sigma(p)$ of (\ref{series1}), interpreted as 
the  CFT correlator at the relevant energy scale,  and $\lambda_{UV}$ is 
given after (\ref{UVlagrangian}). A Fourier transform is performed
by using the properly regulated and normalized 3d Fourier transform of 
$p^{\alpha}$, which is~\cite{gelfand}:  
\begin{equation}
    \int \frac{d^3 p}{2 \pi^2}~ e^{i p\cdot x}~p^{\alpha} = 
     \left( \frac{2}{r} \right)^{3+\alpha} \frac{\Gamma(\frac{\alpha+3}{2})}
     {2 \sqrt{\pi}\; \Gamma(-{\frac{\alpha}{2}})}~.
\end{equation}
After using (\ref{mpuv}), and various gamma-function identities,\footnote{A 
useful identity is: $\Gamma(2 x) = \frac{2^{2 x-1}}{\sqrt{\pi}} \Gamma(x) 
\Gamma(x+\frac{1}{2})$. Note also that a factor of $1/(4 \pi)$ has been 
absorbed in the definition of $V$.} we find {\it precisely} our result from  
the gravity calculation (\ref{grapot1}). 

Consider next the correction, at $r < 1/(k A_1)$,  for the case when the interaction with the CFT is relevant ($\nu_-<0$). Then, we have to sum the chain of bubble graphs as indicated below (recall $\epsilon_\nu = -1$ now):
\begin{eqnarray}
\label{alphaminusrelevant}
V(r) &=& - \mu \frac{\lambda_{UV}^2}{k^2}\int \frac{d^3 p}{2 \pi^2} \; 
\frac{e^{i p\cdot x}}{p^2}\; \left[\epsilon_\nu -  \frac{\lambda_{UV}^2}{k^2}
\frac{\langle {\cal O} {\cal O} \rangle (p)}{p^2}   + \epsilon_\nu 
\frac{\lambda_{UV}^4}{k^4} \left(  \frac{\langle {\cal O} {\cal O}\rangle (p)}
{p^2} \right)^2  -...\right]~,\nonumber \\
&=&  \mu \frac{\lambda_{UV}^2}{k^2}\int \frac{d^3 p}{2\pi^2}\; e^{i p\cdot x}
\frac{1}{p^2 - \frac{\lambda_{UV}^2}{k^2}  \langle {\cal O} {\cal O} 
\rangle (p)} \simeq - \mu \int \frac{d^3 p}{2\pi^2}\; e^{i p\cdot x}  
\frac{1}{\langle {\cal O} {\cal O} \rangle (p)}~,
\end{eqnarray}
where we notice that for the distance scales of interest the CFT correlator 
dominates over $p^2$ in the denominator, as appropriate for a relevant 
coupling. Finally, computing the Fourier transform as before, and using 
(\ref{mpuv}) we again recover precisely the leading term of the potential 
from Eq.~(\ref{grapot2}) from the gravity side.

\subsection{$\alpha_+$ branch holography}

Let us now consider $\nu=\nu_+ = 4 \alpha_+ - 1 > 1$. In this case the 
graviton is always localized on the IR brane. We find, for $A_0 k \gg p \gg A_1 k$, taking the 
upper sign in (\ref{correlator}):
\begin{eqnarray}
\label{apUVcor}
 \Sigma(p)&\simeq& - \left( \frac{M}{k}\right) ^3 k^4 \\
&&\times \left[ (\nu-1) +  q_0^2 \; \frac{1}{4 (\nu-2)}+...
+ c q_0^{[2\nu]-2} + q_0^{2 \nu - 2}  
\frac{\Gamma(2-\nu)}{2^{2 \nu - 2} \Gamma(\nu-1)}\right]\nonumber~,
\end{eqnarray}
indicating that now that the scaling dimension of the operator $\cal O$ is
\begin{equation}
   \Delta_{\cal O} =  \nu+1 = 4 \alpha_+~.
\end{equation}
Thus when $\nu_+>2$ the source coupling to the CFT is irrelevant,
marginal for $\nu_+=2$, while for $1\leq \nu_+< 2$ the coupling is 
relevant. This behavior is plotted in Fig.~\ref{dimOfig}.

At low energies $pL\ll 1$, on the other hand, the series expansion of the correlator is 
given by:\footnote{Note that  (\ref{IRcorr}) does not contradict (\ref{exp2}),
which asserts, instead, that $ \Sigma^{-1}(p)\sim G(p) \simeq 1/p^2$ for 
$p \ll L^{-1}$. In  (\ref{IRcorr}) the pole in $\Sigma$ at $p^2=0$ appears 
only if  the UV cutoff is taken larger than the AdS curvature scale; instead 
in (\ref{exp2}), finite cutoff effects (i.e. mixing with the source) wash 
away the pole. Note that if we formally took  $A_0\rightarrow \infty$ (the 
UV cutoff of the dual theory), after multiplication by $A_0^{2 \nu-2}$ to 
isolate the leading nonanalytic piece in (\ref{apUVcor}), the persistence of 
the pole in (\ref{IRcorr}) would run afoul of the Weinberg-Witten theorem. 
Our dual field theory (\ref{IRlagp}), however, is coupled to gravity  
perturbed by the graviton mass, hence the cutoff cannot be taken large (and 
the source decoupled), due to the strong coupling problem of 
massive gravity~\cite{ags}.}
\begin{equation}
\label{IRcorr}
  \Sigma(p)_{IR} \simeq 
-\left( \frac{M}{k}\right)^3 k^4 \left[ (\nu-1) 
+ q_0^2 \frac{1}{4 (\nu-2)}  - 4\nu(\nu-1)^2 \; 
\frac{A_1^{2\nu}}{A_0^{2\nu}}\; \frac{1}{q_0^2} +\ldots\right]~.
\end{equation}
To obtain (\ref{IRcorr}), we first took the large-$A_0$ limit in 
(\ref{correlator}), obtaining a power series in $q_0$ (of which we   
kept only the leading terms above), with $q_1$-dependent coefficients, and 
subsequently expanded these coefficients for small $q_1$; a more formal way 
of obtaining this is by multiplying the entire correlator by 
$A_0^{2 \nu -2}$ and taking the large $A_0$ limit (see footnote).
Remarkably we see that the correlator now has a pole at $p^2=0$. 
Thus at low energies we can interpret the massless graviton to be  
predominantly a composite of the CFT. This is in contrast to the
$\alpha_-$ branch where no such pole exists. Similar pole structures have
also been identified in Refs.~\cite{pheno2,gp4,cp}.

In addition, we immediately see from (\ref{IRcorr}) that the leading analytic 
piece is a constant, and corresponds to a mass term of order the curvature 
scale for the source field $h_{\mu\nu}$. The interpretation of this fact in 
the dual 4D theory is that the CFT generates a mass for the source so that
it decouples at low energy and the propagating mode is predominantly the 
composite graviton (see also the calculation of the long-distance 
gravitational potential (\ref{alphaposPlanck})). 

The analytic terms of (\ref{IRcorr}) can be used to obtain the long-distance
 Lagrangian:
\begin{equation}
\label{IRlagp}
{\cal L}_{IR}  =   \frac{1}{4}\,h_{\mu\rho} (\Box-m_h^2) h^{\mu\rho} 
+ \frac{\chi}{k}\;  h_{\mu\rho}  T^{\mu\rho}_{CFT}  +
\frac{\chi}{k}   \; h_{\mu\rho}  T^{\mu\rho}_{matter}~ 
+ {\cal L}_{CFT}~, 
\end{equation}
where $\chi = (\nu_+-2)^{1/2} (M/k)^{-3/2}$ and $m_h^2=4(\nu-1)(\nu-2)k^2$. 
If we now write the small-momentum expansion of the  correlator as:
\begin{equation}
    \langle {\cal O}{\cal O}\rangle \simeq \left(M k \right)^3 
    16 \nu_+(\nu_+-1)^2 A_1^{2\nu_+}\frac{1}{p^2}~,
\end{equation}
where $A_0=1$, then the leading contribution to the gravitational potential 
at large distances is given by:
\begin{eqnarray}
\label{alphaposPlanck}
V(r) &\simeq& -\mu \frac{\chi^2}{k^2}\int \frac{d^3 p}
   {2 \pi^2}~e^{i p\cdot x}\;
   \frac{\chi^2}{k^2}\frac{\langle {\cal O} {\cal O} \rangle (p)}
   {m_h^4}~,\nonumber\\
&=& -\mu \frac{\chi^4}{m_h^4} \frac{M^3}{k} 16\nu_+(\nu_+-1)^2 A_1^{2\nu_+}
\int \frac{d^3 p}{2 \pi^2}~ e^{ip\cdot x} \frac{1}{p^2}~,\nonumber\\
&=& -\frac{\mu}{M_P^2 r}~,
\end{eqnarray}
where the source propagator has been approximated by 
$1/(p^2+m_h^2)\simeq 1/m_h^2$  and the Planck mass is given by (using the 
values for $m_h$ and $\chi$ given after (\ref{IRlagp})):
\begin{equation}
     M_P^2 = \left(\frac{M}{k}\right)^3 \frac{k^2}{\nu_+} A_1^{-2\nu_+}
      = \frac{M^3}{k (4\alpha_+-1)}e^{(8\alpha_+-2)\pi k R}~.
\end{equation}
This agrees with the Planck mass formula (\ref{mpuv}) for $\alpha_+>1/2$.  
Note also that further insertions of $\langle {\cal O} {\cal O} \rangle$ in 
(\ref{alphaposPlanck}) are negligible at large distances. 

In fact, the result for the leading long-distance contribution to the  
potential (\ref{alphaposPlanck}) indicates that at $r > L$, we can describe 
long-range physics as due to the exchange of a massless spin-2 field, 
$\bar{h}_{\mu\nu}$---mostly a composite of the CFT---coupled directly to 
matter, thus forgoing the discussion of the massive source in (\ref{IRlagp}).
In this case the Lagrangian is given by
\begin{equation}
\label{IRlagp2}
{\cal L}_{IR} \simeq \frac{1}{4}\left(\frac{M}{k}\right)^3 \frac{k^2}{\nu}\:
A_1^{-2\nu} \;\bar{h}_{\mu\rho}\Box \bar{h}^{\mu\rho} + \bar{h}_{\mu\rho} T^{\mu\rho}_{matter}~,
\end{equation}
where canonically normalising the kinetic term leads to the 
Planck mass (\ref{mpuv}). The IR lagrangian (\ref{IRlagp2}) can, equivalently, be obtained by directly considering  the IR limit
($q_0 \ll1, q_1\ll1$) of the correlator $\Sigma(p)$ of (\ref{correlator}): 
\begin{equation}
\label{IRcorr0}
  \Sigma(p)_{IR} \simeq 
-\left( \frac{M}{k}\right)^3 \frac{k^4}{2} \left[ (A_1^{-2\nu_+}-1)
\; \frac{q_0^2}{2 \nu_+} +...\right]~,
\end{equation}
and interpreting the leading $A_1^{-2\nu_+}$ term as the kinetic term of the 
interpolating long-distance field $\bar{h}_{\mu\rho}$ in (\ref{IRlagp2}).

Continuing on to intermediate energies $L^{-1} \ll p \ll k$, we see immediately
from (\ref{apUVcor}) that there is no longer any pole, since we are now
above the compositeness scale. This is also consistent with the fact that for 
$pL\gg 1$ there is no longer any $1/p^2$ term in the 
Green's function (\ref{exp3}). The analytic terms are identical to those
at low energies, so that the source remains massive. Hence, in this regime 
the dominant contribution to the gravitational potential arises from the CFT, 
which corresponds to the nonanalytic term in (\ref{apUVcor}).

When $\nu_+>2$ the source coupling to the CFT is irrelevant and the 
gravitational potential follows from the coupling to the CFT as depicted in
Fig~\ref{cftfig}. From the analytic terms of (\ref{apUVcor}) we obtain the 
UV Lagrangian:
\begin{equation}
\label{UVlagplus}
{\cal L}_{UV}  =   \frac{1}{4}\,h_{\mu\rho} (\Box-m_h^2) h^{\mu\rho} 
+ \frac{\chi}{k}\;  h_{\mu\rho}  T^{\mu\rho}_{CFT}  +
\frac{\chi}{k}   \; h_{\mu\rho}  T^{\mu\rho}_{matter}~ 
+ {\cal L}_{CFT}~, 
\end{equation}
where $\chi = (\nu_+-2)^{1/2} (M/k)^{-3/2}$ and 
$m_h^2=4(\nu-1)(\nu-2)k^2$. Since there is
no longer any massless pole, the leading contribution to the potential is 
given by:
\begin{eqnarray}
\label{alphaposirrelevant}
V(r) &\simeq& -\mu \frac{\chi^2}{k^2}\int 
    \frac{d^3 p}{2 \pi^2}~e^{i p\cdot x}\; 
   \frac{\chi^2}{k^2}\frac{\langle {\cal O} {\cal O} \rangle (p)}
   {m_h^4}~,\nonumber\\
&=& \mu \frac{\chi^4}{m_h^4}\left(\frac{M}{k}\right)^3
\int \frac{d^3 p}{2 \pi^2}~ e^{ip\cdot x}  
\left(\frac{p}{2k}\right)^{2 \nu -2} \frac{4~\Gamma(2-\nu)}
{\Gamma(\nu-1)}~,
\end{eqnarray}
where  $\langle {\cal O}{\cal O}\rangle$ is 
the nonanalytic part of (\ref{apUVcor}) given by:
\begin{equation}
   \langle {\cal O}{\cal O}\rangle = -\left(\frac{M}{k}\right)^3k^4
   \left(\frac{p}{2k}\right)^{2 \nu -2} 
\frac{4~\Gamma(2-\nu)}{\Gamma(\nu-1)}~.
\end{equation}
Performing the Fourier transform leads precisely to
the result derived purely on the gravity side (\ref{grapot2}).

When $1<\nu_+<2$ the source coupling to the CFT is relevant but the nonanalytic
term is still subdominant compared to the leading mass term in 
(\ref{apUVcor}). In this case no summation is needed beyond the leading 
CFT correction depicted in Fig.~\ref{cftfig} and so the contribution to 
the potential is identical to that obtained in (\ref{alphaposirrelevant}). 
The corresponding Fourier transform then leads to the same expression 
(\ref{grapot2}).

\section{Discussion and conclusion}

We have seen that the graviton zero mode can be smoothly deformed
away from the Planck brane. This deformation requires modifying
the bulk covariant theory at quadratic order by introducing
bulk and boundary mass terms. However, a massless mode only
occurs for a special choice of the bulk and boundary masses. This is an 
additional tuning beyond the usual tuning of bulk and brane cosmological
constants in the RS model. As in the RS case
this tuning may be realized as the result of bulk 
supersymmetry~\cite{gp1,abm}.
At the linearized level there is a 
4D general covariance, which is consistent with the fact that there 
is a massless tensor mode. In addition there is also a massless 
vector mode with a 4D U(1) gauge symmetry. 

However, general relativity is an inherently
nonlinear theory, and it is apparent that higher order nonlinear
interactions will spoil this symmetry. Without further modification
the gravity in our model is different from the full nonlinear Einstein 
theory. This situation may be remedied by modifying
our model at the nonlinear level with the introduction of
nonlinear terms in the bulk and brane, in order to at least 
preserve the 4D general covariance. Thus we expect our
zero mode to remain massless in the nonlinear theory, although this
analysis is beyond the scope of the present paper. 

Nonetheless it is already interesting that a smooth deformation exists 
at quadratic order without the presence of ghosts. In particular the scalar
sector is trivially zero because this is the only
solution consistent with the bulk and boundary equations. Clearly this
is due to the fact that we are working only to quadratic order, and the
scalar modes can possibly appear at the nonlinear level.
Scalar modes may also arise when matter is added on the brane. On the 
phenomenological side, they are certainly needed to reproduce the correct 
gravitational law if the stress--energy tensor of the matter fields is not 
traceless. A similar situation takes place in the usual RS
case. In the compact version with a stabilized radion there are no 
massless scalar excitations. However, a massless scalar mode (most 
easily interpreted as the brane bending mode) arises when matter is present on 
the brane, and allows for the recovery of standard $4$d gravity at 
large scales~\cite{tm}. A similar analysis should also be carried out 
in the set-up we have discussed here.

By the AdS/CFT correpondence there is an interesting 4D dual
interpretation of our model, especially in the case
when the graviton zero mode is localized on the IR brane. This is because 
zero modes localized on the IR brane correspond to CFT bound states
and therefore the dual CFT interpretation would correspond 
to gravity emerging from the strongly coupled gauge theory. In this
model of emergent gravity the UV theory is a gauge (string) theory at the
TeV scale, and the graviton is a composite particle which 
can be associated with the millimeter scale. Thus gravity emerges
as a low energy phenomenon in the IR. This is different from the conventional 
viewpoint that gravity is a fundamental degree of freedom in the UV theory, 
and our model is the first step in constructing and 
understanding this novel possibility.

\section*{Acknowledgments}
We would like to thank Joel Giedt and Alex Pomarol for useful discussions. 
The work of T.G. and M.P. was supported in part by a 
Department of Energy grant 
DE-FG02-94ER40823 at the University of Minnesota. 
T.G. is also supported by a grant from the Office of the 
Dean of the Graduate School of the University of Minnesota, and an 
award from Research Corporation. E.P. acknowledges the support of the 
National Science and Engineering Research Council of Canada. T.G. also
acknowledges the Aspen Center for Physics where part of this work was 
completed.

\appendix 
\def\theequation{\thesection.\arabic{equation}} 
\setcounter{equation}{0}

\section{Bulk gravity action to quadratic order}
\label{appA}

We will present the expansion of the bulk action (\ref{buaz})
around the background RS solution, to quadratic order in the perturbation 
$h_{MN}$, where $g_{MN} = A^2\eta_{MN} + h_{MN}= 
A^2(\eta_{MN}+{\widetilde h}_{MN})$, and $M,N=0,\dots,3,5$.
The Lagrangian density to quadratic order in ${\widetilde h}_{MN}$ is given by:
\begin{eqnarray}
\label{quadraticeinstein}
&&{\cal L}_5[A^2\eta+ h; M]\nonumber\\
 &=& M^3 A^3 \bigg[{\widetilde h}^{MN}
\partial_K\partial^K {\widetilde h}_{MN}-2{\widetilde h}^{MN}\partial_N
\partial^K
{\widetilde h}_{MK}+{\widetilde h}^{MN}\partial_M\partial_N {\widetilde h} 
+ \frac{3}{4}\partial^K {\widetilde h}_{MN}\partial_K {\widetilde h}^{MN}
\nonumber \\
&&+~\partial^N{\widetilde h}\,\partial^M{\widetilde h}_{MN}
-\frac{1}{4}\partial^K{\widetilde h}\,\partial_K {\widetilde h}-\frac{1}{2}
\partial^K
{\widetilde h}^{MN}\partial_M {\widetilde h}_{KN}-
\partial_M{\widetilde h}^{MN}\partial^K
{\widetilde h}_{KN}\nonumber\\ 
&&+~\frac{1}{2}{\widetilde h}\,\partial^M\partial^N{\widetilde h}_{MN}
-\frac{1}{2}{\widetilde h}\,\partial_K\partial^K{\widetilde h}
+\bigg(\frac{A'^2}{A^2}+2\frac{A''}{A}+A^2\frac{\Lambda}{2M^3}\bigg)
\bigg({\widetilde h}_{MN}{\widetilde h}^{MN}-\frac{1}{2}{\widetilde h}^2\bigg)
\nonumber \\
&&+~4\frac{A'}{A}\bigg({\widetilde h}^{MN}{\widetilde h}'_{MN}
-\frac{1}{2}{\widetilde h}\,{\widetilde h}'\bigg)
+12\frac{A'^2}{A^2}\bigg({\widetilde h}^{A5}{\widetilde h}_{A5}-\frac{1}{2}
{\widetilde h}\,{\widetilde h}^{55}\bigg)\nonumber\\
&&-~A^2 \, k^2 (a\,{\tilde h}_{MN} {\tilde h}^{MN} + b\, {\tilde h}^2)\bigg]~.
\label{quadLag}
\end{eqnarray}
where ${\widetilde h} = {\widetilde h}_M^M$, prime ($'$) 
denotes $\partial_5$, and indices are raised and lowered with $\eta_{MN}$.
The last term in (\ref{quadLag}) is the contribution from the bulk mass term.
The corresponding equation of motion arising from ${\cal L}_5$
is given by:
\begin{eqnarray}
&&M^3 A^3\bigg[-\partial^2({\widetilde h}_{MN} - \eta_{MN}{\widetilde h}) 
- \eta_{MN}\partial_A \partial_B {\widetilde h}^{AB}
+ \partial_M \partial^A {\widetilde h}_{AN} + \partial_N \partial^A 
{\widetilde h}_{AM} 
- \partial_M \partial_N {\widetilde h} \nonumber\\
&&-~3\frac{A'}{A}~\bigg({\widetilde h}'_{MN} - \eta_{MN}{\widetilde h}' +
2\eta_{MN}\partial_A {\widetilde h}^{A5} - \partial_M {\widetilde h}_{N5}
- \partial_N {\widetilde h}_{M5}\bigg)
-12\frac{A'^2}{A^2}\eta_{MN} {\widetilde h}_{55}\bigg]\nonumber \\
&&-2A^5(\Lambda+6k^2 M^3) ({\widetilde h}_{MN} - \frac{1}{2} \eta_{MN} 
{\widetilde h})+4 A^5\,k^2\,M^3 (a\,{\widetilde h}_{MN}+
b\,\eta_{MN}\,{\tilde h})= 0~,
\label{5deom}
\end{eqnarray}
where $\partial^2=\Box+\partial_5^2$. Note that 
the R-S solution~\cite{rs} requires that $\Lambda= -6 k^2 M^3$, and
the term involving the cosmological constant in the last line of 
(\ref{5deom}) vanishes. If we consider only the tensor fluctuations 
${\widetilde h}_{MN}={\widehat h}_{\mu\nu}$ as defined in (\ref{metdef})
then we recover the equation of motion (\ref{butmn}).
The remaining equations of motion for the vector (\ref{buv5m}),(\ref{buvmn}) 
and scalar modes (\ref{bus55})-(\ref{busmn2}) follow from the 
$\mu5$ and $55$ components of (\ref{5deom}).
A similar expansion to quadratic order in the metric perturbation, 
but without the bulk mass term, has also been performed in Ref.~\cite{cggp}.

\section{Quadratic action of the graviton zero mode}
\label{appB}

We have seen that by appropriately tuning the bulk and brane mass parameters, 
there is a zero mode both in the tensor and vector sectors. The presence of 
these zero modes signal some symmetries of the starting action. In particular, 
the massless tensor mode signals a 4D general coordinate invariance, while
the massless vector mode signals a 4D gauge invariance. Clearly, these
symmetries only occur at the linearized level in the perturbations. A study 
beyond this order would require considering the inclusion of higher order 
terms in the original action, and it is beyond our current aims. 

It is instructive 
to compute the quadratic action for the tensor zero mode, and to explicitly 
show how the mass term cancels. This calculation also reveals how to 
canonically normalize the graviton in the 4D theory, and therefore how to 
properly define the four dimensional Planck mass. Assuming the Fierz--Pauli
choice $b = - a$, and $\beta_i = - \alpha_i $, with $- \alpha_1 = \alpha_0 
\equiv \alpha$ for the bulk/brane mass terms, respectively, 
the total action up to second order in the tensor perturbations gives:
\begin{eqnarray}
S_{\rm tensor}^{\left( 2 \right)} &=& M^3 \Bigg\{ \int d^5 x \,A^3 \,
\Big[ \sqrt{- {\widehat g}_4} \,{\widehat R}_4 \nonumber\\
&&- \frac{1}{4} {\widehat h}'^{\mu \nu} \, {\widehat h}_{\mu \nu}' - 3 \frac{A'}{A} \, 
{\widehat h}^{\mu \nu} \, {\widehat h}_{\mu \nu}' - \left( 3 \, \frac{A^{'2}}{A^2} 
+ 3 \, k^2 A^2 
+ a \, k^2 A^2 \right) {\widehat h}^{\mu \nu} {\widehat h}_{\mu \nu} \Big] \nonumber\\
&&\pm \int d^4 x \left[ A^4 \, k \left( \frac{3}{2} - \alpha \right) 
{\widehat h}^{\mu \nu} {\widehat h}_{\mu \nu} \right]_i \Bigg\} \,\,,
\label{ac2ten}
\end{eqnarray}
where $+/-$ refers to the UV/IR brane, respectively, and the 
${\widehat h}_{\mu \nu}$ spacetime indices are raised with the (inverse) Minkowski 
metric $\eta^{\mu\nu}$. The first term,
\begin{equation}
\sqrt{- {\widehat g}_4} \, {\widehat R}_4 = \frac{1}{4} \, {\widehat h}^{\mu \nu} \, 
\Box \, {\widehat h}_{\mu \nu} \,\,,
\end{equation}
has the tensorial structure of the quadratic 4D Einstein--Hilbert term
for a transverse--traceless ${\widehat h}_{\mu \nu}$, except that 
${\widehat h}_{\mu \nu}$ 
still depends on the fifth coordinate $z$. One can verify that the 
action~(\ref{ac2ten}) reproduces the tensor mode bulk and brane equations, 
(\ref{butmn}) and~(\ref{junt}), respectively.

To proceed further, we decompose ${\widehat h}_{\mu \nu}$ into the 
eigenmodes (\ref{tbulknomass}) and (\ref{massten}), and integrate over the 
compact coordinate. Hermiticity of the action ensures that eigenmodes with 
different mass eigenvalues are decoupled, so that one is left with an infinite 
sum over the decoupled actions $S_{\rm scalar}^{(2)(n)}$, for each four 
dimensional mode $H_{\mu \nu}^{(n)}$. The last two lines of~(\ref{ac2ten}) 
combine to form the mass term for any given mode. In particular, let us 
consider the zero mode:
\begin{equation}
{\widehat h}_{\mu \nu} = C_1 \, A(z)^{-2(1-\sqrt{1+a})}\,H_{\mu \nu}^{(0)}(x)~,
\end{equation}
as given by (\ref{tbulknomass}), where we only include the part which is 
continuously connected to the RS graviton. In this case, 
the mass terms of (\ref{ac2ten}) combine to give:
\begin{eqnarray}
C_1^2 \, k \, M^3 && \!\!\!\!\!\!\!\!\!\! \Bigg\{ \int_{z_0}^{z_1} d z \, k 
\left[ -2 \, \sqrt{1+a} \, \left( 2+\sqrt{1+a} \right) A 
\left( z \right)^{1+4 \sqrt{1+a} } \right] \nonumber\\
&& \!\!\!\!\!\! + \left( \frac{3}{2} - \alpha \right) 
\left[ A \left( z_0 \right)^{4 \sqrt{1+a}} -  A \left( z_1 \right)^{4 
\sqrt{1+a}} \right] \Bigg\} \,\int d^4 x \, H^{\mu \nu \left( 0 \right)} 
H_{\mu \nu}^{\left( 0 \right)}~,\nonumber\\ 
&& \!\!\!\!\!\!\!\!\!\!\!\!\!\!\!\!\!\!\!\!\!\!\!\!\!\!\!\!\!\!\!\!\!\!\!\!\! 
= C_1^2 \, k \, M^3 \left( \frac{1-\sqrt{1+a}}{2} - \alpha \right) 
\left[ A \left( z_0 \right)^{4 \sqrt{1+a}} -  A \left( z_1 \right)^{4 
\sqrt{1+a}} \right] \int d^4 x \, H^{\mu \nu \left( 0 \right)} 
H_{\mu \nu}^{\left( 0 \right)}.
\end{eqnarray}
Indeed we see that the bulk and brane contributions cancel when 
(\ref{alpha}) is imposed for $\alpha=\alpha_-$, resulting in a massless 
tensor perturbation. Similarly, if we choose the $C_2$ part
of (\ref{tbulknomass}) then we obtain a massless tensor perturbation
for $\alpha=\alpha_+$.

The quadratic 4D action for the zero mode then simply becomes:
\begin{equation}
S_{{\rm scalar}}^{\left( 2 \right) \left( 0 \right)} = \frac{M^3 \, C_1^2}
{2 k\left( 4 \alpha - 1 \right)} \left[
A(z_1)^{2( 1-4 \alpha)} - A(z_0)^{2 (1- 4 \alpha)} \right] 
\int d^4 x \sqrt{-g_4} \, R_4~,
\end{equation}
where $g_4 \,,\, R_4$ now refer to the standard 4D metric $g_{\mu \nu,4} 
= \eta_{\mu \nu} + H_{\mu \nu}^{(0)}(x)$. By adding a source term on the 
brane, it is straightforward to see that the coupling of $H_{\mu \nu}^{(0)}$ 
to brane fields is set by the 4D Planck mass~(\ref{mpl}).

\section{Quadratic action of the scalar modes}
\label{appC}

The equations for the scalar perturbations can be directly obtained from the 
second order action in the scalar perturbation. The derivation of this
action is quite involved (we extend the computation of Ref.~\cite{kmp}, 
performed for the covariant case), but the result is relatively simple. The
computation of the action supports the choice of the (generalized) 
Fierz--Pauli mass term for the perturbations. Indeed, the expansion of the 
mass terms gives the following higher derivative kinetic terms:
\begin{equation}
S_{\rm scalar}^{\left( 2 \right)} \supset 4 \, k^2 \int d^5 x\,
A^5\left( z \right) \left( a + b \right) \left( \Box E \right)^2 
+ \sum_i 4 \, k \int d^4 x \,A^4(z_i)(\alpha_i + \beta_i) 
\left( \Box E\right)_i^2 \,\,,
\end{equation}
where the sum over $i$ refers to the two boundary branes.
As in the 4D case, the Fierz--Pauli choice ($a+b = \alpha_i + \beta_i = 0 $) 
eliminates these pathological higher derivative terms from the 
action~\cite{ags}. Employing this choice, and relating the brane mass 
coefficients as $\alpha_{0,1}= \pm \alpha \,$, the action for the scalar 
perturbations is:
\begin{eqnarray}
S_{\rm scalar}^{\left( 2 \right)} &=& \int d^5 x \, A^3 
\nonumber\\
&&\times\Bigg\{- 6 {\widehat \psi} \Box \left( {\widehat \psi} + {\widehat \phi} \right) 
+ 12 {\widehat \psi}^{' 2} + 24 \frac{A'}{A} {\widehat \psi}'
\left( 2 \, {\widehat \psi} - {\widehat \phi} \right) + 12 \frac{A^{' 2}}{A^2} 
\left({\widehat \phi}^{2} + 8 \, {\widehat \psi}^2 \right) \nonumber\\
&& + \,a \, k^2 A^2 \left[ 32 \, \phi \, \psi + 48 \, \psi^2 
+ 24 \, \psi \, \Box E + 8 \, \phi \Box E + 2 \, B \Box B \right] 
\Bigg\}~, \nonumber\\
&\pm& \int d^4 x \, A^3 \Bigg\{ 24 \, \frac{A'}{A} \, {\widehat \psi}^2 
+ 6 {\widehat \psi} \Box {\widehat \zeta} + 3 \, \frac{A'}{A} \, 
{\widehat \zeta} \Box {\widehat \zeta} \nonumber\\
&&\quad\quad\quad+ \,\alpha \, k \, A \left[ 48 \left( \psi + \frac{A'}{A} 
\zeta \right)^2 + 24 \left( \psi + \frac{A'}{A} \zeta \right) \Box E 
\right] \Bigg\}_i~,
\label{ac2sca}
\end{eqnarray}
where $+/-$ refers to the UV/IR brane, respectively. Note that only gauge 
invariant combinations ${\widehat \psi}, \, {\widehat \phi}, 
\, {\rm and} \, {\widehat \zeta}$ appear when there are no bulk or boundary 
masses. This is due to the fact that the usual 
action for gravity is general coordinate invariant. This symmetry has been 
made manifest in (\ref{ac2sca}), by rewriting a total derivative in 
the bulk as a boundary term, which then produces a sum of two brane terms. 
For instance, this is the origin of an $E''$ term which is not present 
in the original action, but which is needed to produce the 
${\widehat \psi}^{'2}$ term in the bulk. Clearly, this procedure does not 
change the action, but allows one to write it in a more compact and manifestly 
covariant form. The terms proportional to the bulk/brane masses are instead 
not general coordinate invariant, and for this reason they cannot be 
rewritten in terms of gauge invariant quantities only.

To obtain the equations of motion from~(\ref{ac2sca}), it is convenient 
to work with the original total derivative bulk terms, and compute the 
Euler-Lagrange equations by the usual procedure. In fact the simplest way
to proceed is to rewrite (\ref{ac2sca}) as a bulk action by promoting 
$\zeta$ to a bulk field, which evaluates to $\zeta_i$ at the two boundaries,
and then vary this 5D action. In general when we vary terms containing 
$f'$ (where $f$ denotes any of the scalar perturbations), we produce 
terms proportional to $\delta f'$. These terms are dealt with by integrating 
by parts the variation of the action:
\begin{equation}
\int d^5 x \, \delta f' \, \left[ \dots \right] = \int d^5 x 
  \left\{ \delta f \, \left[ \dots \right] \right\}' 
   - \int d^5 x \, \delta f \,  \left[ \dots \right]'~.
\label{fvar}
\end{equation}
The last term in (\ref{fvar}) enters in the usual Euler--Lagrange equations, 
while the total derivative is usually assumed to vanish when evaluated at 
infinity. However, in the presence 
of branes, this term must also separately vanish on shell, for each brane.
It is precisely this requirement that leads to the boundary conditions for 
the bulk fields.

Using this procedure, the bulk/brane equations of motion can be readily 
computed. There are four bulk equations:
\begin{eqnarray}
&& \Box \left[ {\widehat \psi}' - \frac{A'}{A} {\widehat \phi} - \frac{2}{3} 
\, a \, k^2 \, A^2 \, B \right] = 0~, \nonumber\\
&& \Box \left[ {\widehat \psi}'' - \frac{A'}{A} {\widehat \phi}' 
+ 3 \frac{A'}{A} {\widehat \psi'} - 4 \frac{A^{'2}}{A^2} {\widehat \phi} 
- \frac{4}{3} \, a \, k^2 A^2 \left( 3 \, \psi + \phi \right) \right] = 0~, 
\nonumber\\
&& \frac{A'}{A}\left({\widehat \psi}' - \frac{A'}{A} {\widehat \phi} \right) 
- \frac{4}{3} \, a \, k^2 \, A^2 \, \psi + \Box \left[ \frac{1}{4} 
{\widehat \psi} - \frac{a}{3} \, k^2 \, A^2 \, E \right] = 0~, \nonumber\\
&& {\widehat \psi}'' - \frac{A'}{A} {\widehat \phi}' + 3 \frac{A'}{A} 
{\widehat \psi'} - 4 \frac{A^{'2}}{A^2} {\widehat \phi} - \frac{4}{3} \, 
a \, k^2 A^2 \left( 3 \, \psi + \phi \right)~\nonumber\\
&& + \,\Box \left[ \frac{1}{2} {\widehat \psi} + \frac{1}{4} {\widehat \phi} 
- a \, k^2 \, A^2 \, E \right] = 0~,
\label{bulksca2}
\end{eqnarray}
which are obtained by varying $B \,, E \,, \phi \,,$ and $\psi \,$, 
respectively (since $\zeta$ is defined only on the two boundaries).

For the boundary conditions one would naively expect five equations. 
However, one can verify that once the boundary term in~(\ref{ac2sca}) 
is rewritten as a bulk term, $\phi'$ and $B'$ do not appear in the total 
action. Hence, there are no boundary conditions arising from the variation of 
the action with respect to $\phi$ and $B$. This leaves only three 
boundary conditions which follow from varying $E \,, \psi \,,$ 
and $\zeta$ in the action~(\ref{ac2sca}). These are respectively given
by the following equations evaluated on the two branes:
\begin{eqnarray}
&&\!\!\!\! \Box \left[ \psi' - \frac{A'}{A} \phi - 4 \, \alpha \, k \, A 
\left( \psi + \frac{A'}{A} \, \zeta \right) \right] = 0~, \label{j2e} \\
&& \!\!\!\! \psi' - \frac{A'}{A} \, \phi  - 4 \, \alpha \, k \, A 
\left( \psi + \frac{A'}{A} \, \zeta \right) + \frac{1}{4}\Box 
\left[ E' - B - \zeta - 4\,\alpha \, k \, A \, E \right] = 0~, \label{j2p} \\
&& \!\!\!\! 4 \, \alpha \, k \frac{A'}{A} \left( \psi + \frac{A'}{A} \, 
\zeta \right) + \frac{\Box}{A} \left[ \alpha \, k \, A' \, E + \frac{1}{4} 
\left( \psi + \frac{A'}{A} \, \zeta \right) \right] = 0~. \label{jz2}
\end{eqnarray}

For the massless modes (i.e, imposing $\Box \equiv 0$) the solution
of the above equations is $\psi = \phi = \zeta = 0 \,$. The remaining 
combination $B-E'$ is undetermined and one can verify that when the other 
modes are absent, this combination gives a vanishing contribution to the 
action~(\ref{ac2sca}). In the covariant case, we have seen that out of the 
original scalar modes $\psi ,\, \phi ,\, E ,\, B ,\, \zeta$ 
only ${\widehat \psi} ,\, {\widehat \phi} ,\, {\widehat \zeta}$ appear in the quadratic 
action for the perturbations. This is a consequence of 5D general coordinate
invariance, which guarantees that two modes are not present in the action 
at all orders. In the present case, higher order terms could make the 
remaining modes dynamical. However, to study in a consistent way the 
non--covariant theory beyond quadratic order in the action requires 
introducing higher-order terms in addition to the quadratic bulk/brane mass 
terms~(\ref{buaz}) and ~(\ref{fpbound}). This is beyond the scope of the 
present analysis. 

For the massive modes, one can check that the bulk equations~(\ref{bulksca2}) 
are equivalent to the equations (\ref{bus55}) - (\ref{busmn2}). Also, the 
boundary equations~(\ref{j2e}) and~(\ref{j2p}) are equivalent (for the 
massive modes) to the two junction conditions (\ref{juns1}), and 
(\ref{juns2}). However Eq.~(\ref{jz2}) is a third independent boundary 
condition, which as discussed in the main text, cannot be obtained through 
the usual junction/Israel condition procedure. Eliminating $\zeta$ 
in~(\ref{jz2}) through the other two boundary conditions gives:
\begin{equation}
4\,\frac{A'}{A} \left( {\widehat \psi}' - \frac{A'}{A} \, {\widehat \phi} \right)
+ \Box {\widehat \psi} = 0~.
\end{equation}
Using the third equation of~(\ref{bulksca2}), this equation simplifies
to:\footnote{Note that there is no problem with combining bulk and boundary 
equations. Indeed, the value of any function $f(z_i)$ 
in the boundary equations is defined as lim $f(z)$ for 
$z \rightarrow z_i$ in the fundamental domain.}
\begin{equation}
    a \left( 4 \, \psi + \Box E \right) = 0~.
\label{simp3eqn}
\end{equation}
This boundary condition vanishes identically when $a=0$. This shows that 
the ``additional'' boundary condition~(\ref{jz2}) is redundant in 
the standard case without bulk or boundary mass terms. Instead when 
$a\neq 0$ we obtain an independent boundary condition, which appears as 
Eq.~(\ref{juns3}) in Section 3.1.

\section{Derivation of the gravitational potential}
\label{appD}

We will now derive the leading terms in the 
expressions~(\ref{grapot1})-(\ref{grapot2}). We start from the action for 
the Kaluza-Klein modes of the $5$d tensor mode, coupled to a conserved 
matter source on the UV brane. The action can be written in the form:
\begin{equation}
S = \sum_n c_n \int d^4 x~ H^{(n)\mu \nu} \left( \Box - m_n^2 \right)
H_{\mu \nu}^{(n)} + \frac{1}{2}Z_n(0) \int d^4 x~H^{(n)}_{\mu \nu} 
T^{\mu \nu}~,
\end{equation}
where $T_{\mu \nu}$ denotes the energy-momentum tensor of
the matter source, and the variables of the 
tensor mode wavefunctions are separated as:
\begin{equation}
{\widehat h}_{\mu \nu}(x,z) = \sum_n H_{\mu \nu}^{(n)} \left( x \right) Z_n
\left( z \right)~.
\end{equation}
The coefficients $c_n$ are given by:
\begin{equation}
\label{cen}
c_n = \frac{M^3}{4} \int_0^{z_1} d z \, A^3 \, Z_n^2(z)~.
\end{equation}

Correspondingly, the gravitational potential generated by a static
mass $\mu$ on the UV brane, measured at the distance $r$ from $\mu
\,$, is given by:
\begin{equation}
V = - \frac{\mu}{8} \sum_n \frac{Z_n^2 \left( 0 \right)}{c_n} \,
\frac{{\rm e}^{- m_n \, r}}{r} \,\,.
\label{summodes}
\end{equation}

All the modes mediate a (gravitational) attraction. The
sum~(\ref{summodes}) includes the long range contribution from the
zero mode, giving rise to the standard Newtonian potential,
characterized by the Planck mass, $M_P$~(\ref{mpuv}), while 
each Kaluza-Klein mode $n$ provides a Yukawa-type force contribution, which
is relevant at $r \lappeq m_n^{-1} \,$. We are interested in computing 
the gravitational potential at the ``intermediate'' 
distances~(\ref{intermediate}). For these distances, the largest 
contribution to the potential~(\ref{summodes}) is given by modes with 
mass $m$ satisfying:
\begin{equation}
A_1 \, k \ll m \ll k \,\,.
\label{range}
\end{equation}
We can use this condition in the expansion of the Bessel functions
which characterize the bulk profile of the Kaluza-Klein modes~(\ref{massten}).
Moreover, the mass spectrum in this range is given in Eq.~(\ref{tenKKmass}).

The ``plus'' and ``minus'' branches can be discussed simultaneously by 
introducing the parameters $\nu_\pm \equiv \pm \left( 4 \alpha_\pm - 1 \right) 
\,$. They satisfy $2 \sqrt{1+a} = \nu_\pm \mp 1\,$, so that $\nu_\pm$ 
ranges from $\pm1$ to $+\infty \,$, as $a$ ranges from $-1$ to $+ \infty \,$ 
(the RS point is at $\nu_- = 1\,$). We can set $C_1 = 1\,$ in the mode 
solutions~(\ref{massten}), since the normalization cancels in~(\ref{summodes})
(also, for shorthand we write $\nu$ rather than $\nu_\pm$). We then have:
\begin{equation}
Z_n \left( z \right) = A^{-2} \left[ J_{\nu \mp 1} \left( \frac{m_n}{k A}
\right) + C_2 \, Y_{\nu \mp 1} \left( \frac{m_n}{k A} \right) \right]
\;\;,\;\; C_2 = - \frac{J_\nu \left( \frac{m_n}{k} \right)}{Y_\nu \left(
\frac{m_n}{k} \right)} = - \frac{J_\nu \left( \frac{m_n}{k A_1}
\right)}{Y_\nu \left( \frac{m_n}{k A_1} \right)} \,\,.
\end{equation}
which give the following exact expressions:
\begin{equation}
Z_n \left( 0 \right) = \mp \frac{2 \, k}{m_n \, \pi \, Y_\nu \left(
\frac{m_n}{k} \right)} \;\;,\;\; c_n = \frac{M^3 \, k}{2 \, \pi^2 \,
m_n^2} \left[ \frac{1}{Y_\nu^2 \left( \frac{m_n}{k A_1} \right)} -
\frac{1}{Y_\nu^2 \left( \frac{m_n}{k} \right)} \right]~.
\label{exactexp}
\end{equation}
Accordingly, we obtain:
\begin{equation}
\frac{Z_n^2 \left( 0 \right)}{c_n} = \frac{8 \, k}{M^3} \,
\frac{Y_\nu^2 \left( \frac{m_n}{k A_1} \right)}{Y_\nu^2 \left(
\frac{m_n}{k} \right) - Y_\nu^2 \left( \frac{m_n}{k A_1} \right)}
\simeq \frac{8 \, k}{M^3} \, \frac{Y_\nu^2 \left( \frac{m_n}{k A_1}
\right)}{Y_\nu^2 \left( \frac{m_n}{k} \right)}~.
\end{equation}

Inserting this ratio back into the potential~(\ref{summodes}), and 
taking the expression~(\ref{mpuv}) for the 4D Planck mass on the UV brane, 
we obtain,
\begin{equation}
V \simeq - \frac{\mu}{M_P^2 \, r} \, \left[ 1 + \frac{k \, M_P^2}{M^3} \,
\sum_{n > 0} \, \frac{Y_\nu^2 \left( \frac{m_n}{k A_1}
\right)}{Y_\nu^2 \left( \frac{m_n}{k} \right)} \, {\rm e}^{-m_n \, r} \right] \,\,.
\label{short}
\end{equation}

In the range~(\ref{range}), the sum becomes:
\begin{eqnarray}
\label{Potential}
&&\sum\limits_{n>0} 
\frac{Y_\nu^2 \left( \frac{m_n}{k A_1}
\right)}{Y_\nu^2 \left( \frac{m_n}{k} \right)} \; {\rm e}^{- m_n \, r} 
\nonumber \\
 &\simeq& \frac{\pi A_1}{2^{2|\nu|} \Gamma(|\nu|)^2 } \sum\limits_{n>0}^{A_1^{-1}}
\left(\frac{m_n}{k}\right)^{2 |\nu| - 1} \, {\rm sin}^2 \left( \frac{m_n}{A_1 \, k} - \frac{\pi \vert \nu \vert}{2} - \frac{\pi}{4} \right) \, {\rm e}^{- m_n \, r} \nonumber \\
&\simeq&  \frac{\pi A_1}{2^{2|\nu|-1} \Gamma(|\nu|)^2 } \sum\limits_{n>0}^{A_1^{-1}}
\left(\frac{m_n}{k}\right)^{2 |\nu| - 1}  \, {\rm e}^{- m_n \, r} \nonumber \\
&\simeq& \frac{1}{2^{2|\nu|-1} \Gamma(|\nu|)^2 } \int\limits_{0}^{\infty} d n\; n^{2|\nu|-1} \, {\rm e}^{- n k r} = \frac{\Gamma(2 |\nu|)}{2^{2|\nu|-1}
\Gamma(|\nu|)^2 }\frac{1}{(k r)^{2 |\nu|}}~,
\end{eqnarray}
where we have first used the result~(\ref{massten}), and then we have approximated the expression for the masses as $m_n \simeq \pi k A_1 n$ in the relevant regime. The potential~(\ref{short}) then rewrites:
\begin{equation}
\label{potential1}
V \simeq - \frac{\mu}{M_{P}^2 r} \left[ 1 + \frac{k \, M_P^2}{M^3} \, 
\frac{1}{(k r)^{2 |\nu_\pm|}} \frac{\Gamma(2 |\nu_\pm|)}{2^{2|\nu_\pm|-1} 
\Gamma(|\nu_\pm|)^2 }\right]~.
\end{equation}
where we have reintroduced the suffix $\pm$ which characterizes the two 
branches.

The ratio $k \, M_P^2 / M^3$ controls the relative contribution of the zero mode and the Kaluza-Klein tower. For $\alpha < 1/4 \,$, the zero mode is more localized towards the UV brane, and the massive Kaluza-Klein modes have a negligible effect in the regime~(\ref{intermediate}). Substituting the value for $M_P$ given in eq.~(\ref{mpuv}), and identifying $\nu_- = 1- 4 \alpha \equiv \xi \,$, we obtain the result~(\ref{grapot1}) of the main text. In the complementary interval $\alpha > 1/4 \,$, the zero mode is localized towards the IR brane, and the relative contribution from the Kaluza-Klein modes significantly increases. This interval is covered both by $\nu_- < 0 \,$, and $\nu_+ \,$. We can describe it by a unique parameter $\xi \equiv 4 \alpha - 1 \,$, ranging from zero to infinity, which is identified with $- \nu_-$ from $0$ to $1$, and with $\nu_+$ from $1$ to $\infty$. This leads to the expression~(\ref{grapot2}) given in the main text.

\end{document}